\documentclass[twocolumn]{aastex61}
\usepackage{natbib}
\usepackage{graphics}

\accepted{ApJ; September 14, 2018}

\shorttitle{ALMA observations of high-z starbursts}
\shortauthors{Silverman et al.}

\begin{document}


\title{The molecular gas content and fuel efficiency of starbursts at $z\sim1.6$ with ALMA}

\author{J.~D.~Silverman}
\email{silverman@ipmu.jp}
\affiliation{Kavli Institute for the Physics and Mathematics of the Universe, The University of Tokyo, Kashiwa, Japan 277-8583 (Kavli IPMU, WPI)}

\author{W. Rujopakarn}
\affiliation{Department of Physics, Faculty of Science, Chulalongkorn University, 254 Phayathai Road, Pathumwan, Bangkok 10330, Thailand}
\affiliation{National Astronomical Research Institute of Thailand (Public Organization), Don Kaeo, Mae Rim, Chiang Mai 50180, Thailand}
\affiliation{Kavli Institute for the Physics and Mathematics of the Universe, The University of Tokyo, Kashiwa, Japan 277-8583 (Kavli IPMU, WPI)}

\author{E.~Daddi}
\affiliation{Laboratoire AIM, CEA/DSM-CNRS-Universite Paris Diderot, Irfu/Service d'Astrophysique, CEA Saclay}

\author{A. Renzini}
\affiliation{Istituto Nazionale de Astrofisica, Osservatorio Astronomico di Padova, v.co dell'Osservatorio 5, I-35122, Padova, Italy}

\author{G.~Rodighiero}
\affiliation{Dipartimento di Fisica e Astronomia, Universita di Padova, vicolo Osservatorio, 3, 35122, Padova, Italy}

\author{D. Liu}
\affiliation{Laboratoire AIM, CEA/DSM-CNRS-Universite Paris Diderot, Irfu/Service d'Astrophysique, CEA Saclay}

\author{A.~Puglisi}
\affiliation{Laboratoire AIM, CEA/DSM-CNRS-Universite Paris Diderot, Irfu/Service d'Astrophysique, CEA Saclay}
\affiliation{Istituto Nazionale de Astrofisica, Osservatorio Astronomico di Padova, v.co dell'Osservatorio 5, I-35122, Padova, Italy}

\author{M.~Sargent}
\affiliation{Astronomy Centre, Department of Physics and Astronomy, University of Sussex, Brighton, BN1 9QH, UK}

\author{C. Mancini}
\affiliation{Dipartimento di Fisica e Astronomia, Universita di Padova, vicolo Osservatorio, 3, 35122, Padova, Italy}
\affiliation{Instituto Nazionale de Astrofisica, Osservatorio Astronomico di Padova, v.co dell'Osservatorio 5, I-35122, Padova, Italy}

\author{J. Kartaltepe}
\affiliation{School of Physics and Astronomy, Rochester Institute of Technology, 84 Lomb Memorial Drive, Rochester, NY 14623, USA}

\author{D. Kashino}
\affiliation{Department of Physics, ETH Z{\" u}rich, Wolfgang-Pauli-Strasse 27, CH-8093, Z{\" u}rich, Switzerland}

\author{A. Koekemoer}
\affiliation{Space Telescope Science Institute, 3700 San Martin Drive, Baltimore, MD, 21218, USA}

\author{N. Arimoto}
\affiliation{Astronomy Program, Department of Physics and Astronomy, Seoul National University, 599 Gwanak-ro, Gwanak-gu, Seoul, 151-742, Korea}
\affiliation{Subaru Telescope, 650 North A'ohoku Place, Hilo, Hawaii, 96720, USA}

\author{M. B\'{e}thermin}
\affiliation{European Southern Observatory, Karl-Schwarzschild-Str. 2, 85748 Garching, Germany}

\author{S. Jin}
\affiliation{Laboratoire AIM, CEA/DSM-CNRS-Universite Paris Diderot, Irfu/Service d'Astrophysique, CEA Saclay}
\affiliation{Key Laboratory of Modern Astronomy and Astrophysics in Ministry of Education, School of Astronomy and Space Sciences, Nanjing University, Nanjing, 210093}

\author{G. Magdis}
\affiliation{Cosmic DAWN Centre, Niels Bohr Institute, University of Copenhagen, Juliane Mariesvej 30, 2100, Copenhagen, Denmark}

\author{T. Nagao}
\affiliation{Research Center for Space and Cosmic Evolution, Ehime University, 2-5 Bunkyo-cho, Matsuyama 790-8577, Japan}

\author{M. Onodera}
\affiliation{Subaru Telescope, 650 North A'ohoku Place, Hilo, Hawaii, 96720, USA}

\author{D. Sanders}
\affiliation{Institute for Astronomy, University of Hawaii, 2680 Woodlawn Drive, Honolulu, HI, 96822, USA}

\author{F. Valentino}
\affiliation{Dark Cosmology Centre, Niels Bohr Institute, University of Copenhagen, Juliane Maries Vej 30, DK-2100 Copenhagen, Denmark}



\begin{abstract}

We present an analysis of the molecular gas properties, based on CO(2 - 1) emission, of twelve starburst galaxies at $z\sim1.6$ selected by having a boost ($\gtrsim4\times$) in their star formation rate (SFR) above the average star-forming galaxy at an equivalent stellar mass. ALMA observations are acquired of six additional galaxies than previously reported through our effort. As a result of the larger statistical sample, we significantly detect, for the first time at high-z, a systematically lower $L^{\prime}_{CO}$/$L_{IR}$ ratio in galaxies lying above the star-forming `main sequence' (MS). Based on an estimate of $\alpha_{CO}$ (i.e., the ratio of molecular gas mass to $L^{\prime}_{CO~1-0}$), we convert the observational quantities (e.g., $L^{\prime}_{CO}$/$L_{IR}$) to physical units ($M_{gas}$/$SFR$) that represent the gas depletion time or its inverse, the star formation efficiency. We interpret the results as indicative of the star formation efficiency increasing in a continuous fashion from the MS to the starburst regime, whereas the gas fractions remain comparable to those of MS galaxies. Although, the balance between an increase in star-formation efficiency or gas fraction depends on the adopted value of $\alpha_{CO}$ as discussed. 

\end{abstract}





\section{Introduction}

On occasion, galaxies experience a rapid rise in their production rate of forming new stars, usually referred to as a starburst \citep[e.g.,][]{Sanders1996}. Such explosive phenomena, in the local Universe, are the result of the merger of two massive, gas-rich galaxies that induce intense star formation and can boost the growth of their central supermassive black hole \citep[e.g.,][]{DiMatteo2005,Hopkins2006,Volonteri2015}. It remains to be demonstrated whether the same process is also responsible for starburst galaxies in the early Universe, with star formation rates (SFR) several times higher than those of more common galaxies. Recent hydrodynamical simulations of galaxy mergers \citep{Fensch2017} at high redshifts show rather mild boosts in star formation, as compared to the typical star-forming galaxy population, that is attributed to the already enhanced SFRs as a result of their high gas fractions. While such starburst galaxies do not appear to be the leading mechanism responsible for the cosmic history of star formation \citep{Rodighiero2011,Sargent2012,Lackner2014}, they may still represent a crucial passage in the life cycle of galaxies. 

Local starbursts appear to support the scenario of having a higher efficiency in forming stars \citep{Solomon1997} and suggest a different ÒmodeÓ of star formation distinct from that of typical star-forming galaxies \citep{Daddi2010b,Genzel2010}. However, it is still unclear whether starbursts, including those at higher redshifts, are the result of either higher levels of molecular gas out of which stars form (i.e., a higher gas fraction; \citealt{Scoville2016,Lee2017}), a higher efficiency to form stars from a given supply of gas \citep{Magdis2012a,Genzel2015,Silverman2015a,Elbaz2017} or a combination of both \citep{Combes2013,Scoville2017,Tacconi2018}. It is therefore imperative to establish whether such different modes of star formation are active at earlier stages in the evolution of our Universe when most of the star formation was actually taking place. Alternatively, the efficiency to form stars may be a continuous function with distance above the star-forming 'main-sequence' \citep[MS hereafter;][]{Speagle2014,Renzini2015}. With sufficient investigations, we can also determine whether mergers are the sole mechanism responsible for enhanced star formation efficiency or can other processes, such as violent disk instabilities \citep{Bournaud2010,DekelBurkert2014}, drive star formation rates well above that of the typical star-forming MS. \citet{Carilli2013} present a overview of the molecular gas properties of high redshift star-forming galaxies as related to this topic.

To address these issues, we have undertaken a study, as first presented in \citet{Silverman2015a}, to measure the molecular gas properties using carbon monoxide $^{12}$CO, primarily the J = 2 - 1 transition, of galaxies in the COSMOS field \citep{Scoville2007} having star formation rates well-above ($\gtrsim4\times$) the star-forming MS at $z\sim1.6$. Observations are mainly acquired with the Atacama Large Millimeter/submillimeter Array (ALMA) and supplemented with those from the Northern Extended Millimeter Array (NOEMA) Interferometer. The rotational transition CO(2 - 1), is at an excitation level close to the locally-calibrated tracer CO(1 - 0) of the total molecular gas mass, thus has been used for many studies of high-redshift galaxies to date \citep[e.g.,][]{Tacconi2018}. 

Our starburst sample has been selected to span a narrow range of accurate spectroscopic redshifts ($z\sim1.6$) from which it has been drawn \citep{Silverman2015b}. At these redshifts, there are still limited numbers of starburst galaxies with CO observations in the literature \citep[see Figure 2 of ][]{Tacconi2018}. The stellar mass and SFRs of galaxies in our sample are based on mid- and far-IR observations from the $Spitzer$ and $Herschel$ satellites, placing them securely in the starburst class \citep{Rodighiero2011} and matching fairly well in stellar mass to existing samples of more normal galaxies with CO detections at a similar redshift. Being located in the COSMOS field, we have a wealth of multi-wavelength data across the electromagnetic spectrum thus our approach, in particular the selection method, complements that used in the literature for starburst galaxies, such as the brightest sub-mm galaxies (SMG) at high-z \citep[e.g.,][]{Ivison2005,Ivison2011,Casey2011,Casey2017,Danielson2017}. Compared to classical SMGs, our selection avoids contamination by massive galaxies on the MS \citep{Magnelli2012}, reduces SED bias compared to SMGs selection (which prefers cold objects), and is able to pick objects even with less extreme SFRs, but much higher above the MS population. We refer the reader to \citet{Casey2014} that provides a comprehensive review of the vast literature on dust-obscured star-forming galaxies at high redshift.

To build upon our past study, we report here on ALMA observations of the molecular gas content as traced by CO(2 - 1) of six additional starburst galaxies that have SFRs highly elevated ($\gtrsim4\times$) from the star-forming MS at $z\sim1.6$. This larger sample of twelve galaxies enables us to confirm trends (e.g., $L^\prime_{\rm CO}/<L^\prime_{\rm CO}>_{\rm MS}$ vs. $\delta _{MS}=sSFR /<sSFR>_{\rm MS}$, where $sSFR=SFR/M_{stellar}$) seen in the previous study with higher statistical significance due to the larger size of the sample and higher boost in SFR above the star-forming MS. We present our results with consideration of the likely range of the factor ($\alpha_{CO}$) required to convert CO luminosity to molecular gas mass \citep{Bolatto2013,Carleton2017}, a critical aspect of this analysis. Correspondingly, we have measured $\alpha_{CO}$ based on dynamical arguments for three of our sources, as presented in \citet{Silverman2015a} and Silverman et al. (2018, in press), and have information on the metallicity from the [NII]/H$\alpha$ ratio acquired through near-infrared spectroscopy \citep{Zahid2014,Puglisi2017,Kashino2017}. Throughout this work, we refer to the total molecular gas mass as primarily composed of H$_2$ and He, and assume $H_0=70 $ km s$^{-1}$ Mpc$^{-1}$, $\Omega_{\Lambda}=0.7$, $\Omega_{\rm{M}}=0.3$. We use stellar masses and SFRs converted to a scale based on a Chabrier initial mass function (IMF); we chose this IMF (different than presented in \citet{Silverman2015a}) to aid in comparisons with the literature \citep{Sargent2014,Tacconi2018}.
       




\section{Selection of the sample and physical characteristics}

\begin{figure}
\includegraphics[width=9cm]{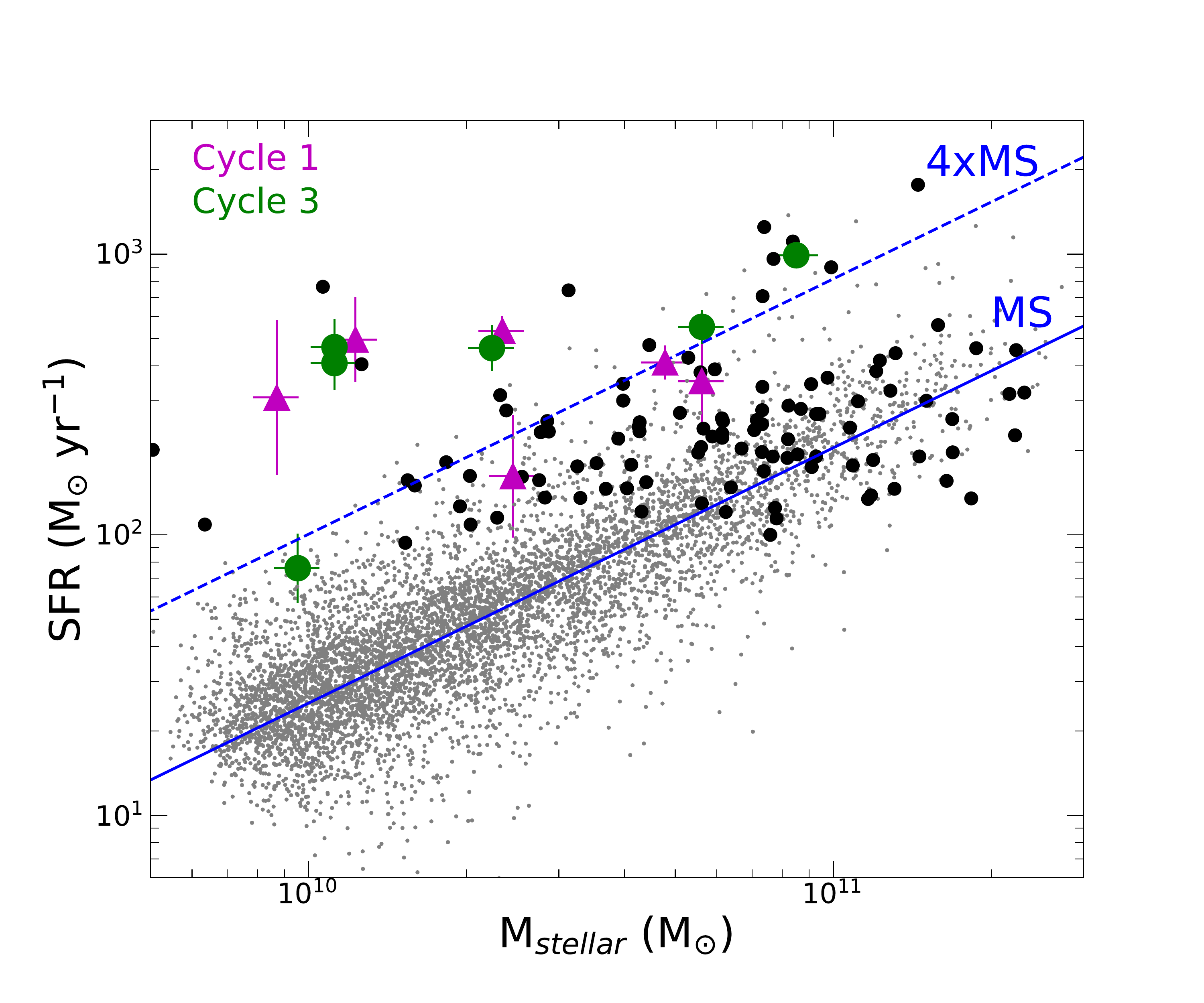}
\caption{SFR versus M$_{stellar}$ for star-forming galaxies (including starbursts detected by $Herschel$) at $1.4<z<1.7$. The large colored symbols indicate the twelve starbursts observed by ALMA (Cycle 1: magenta, Cycle 3: green) that constitute the focus of our study. With black symbols, we indicate the location of $Herschel$-detected galaxies that have spectroscopic redshifts over the same interval from the FMOS-COSMOS survey. Smaller circles show the location of MS galaxies (grey) having photometric redshifts from \citet{Laigle2016}. The solid line indicates the MS, as determined by a fit to the grey data points having sSFR $>$ 2$\times$10$^{-10}$ yr$^{-1}$, and a parallel track (dashed) at an elevated rate of 4$\times$ above the MS that illustrates the typical boosts in SFR for our starbursts.} 
\label{fig:selection}
\end{figure}

To construct a robust sample of starburst galaxies at $z\sim1.6$, we first acquired spectroscopic redshifts of $Herschel$ sources through the FMOS-COSMOS survey \citep{Silverman2015b}, a spectroscopic survey of star-forming galaxies at $1.4 \lesssim z \lesssim 1.7$ in the near-infrared (1.11 - 1.35 and 1.6 - 1.8 $\mu$m) using the instrument FMOS \citep{Kimura2010} on the Subaru Telescope. Fibers were allocated to $Herschel$ sources with a high priority, especially for those having SFRs that place them above ($>4\times$) the MS (see below). Spectroscopic redshifts are determined from the centroid of the H$\alpha$ emission line with additional lines ([NII]$\lambda$6584, [OIII]$\lambda5008$ and H$\beta$) providing further assurance of the redshift. These emission lines also provide a measure of the level of dust extinction and chemical enrichment. On the rest-frame optical emission-line properties of high-z starbursts, we refer to a separate study \citep{Puglisi2017} that makes use of a larger sample than presented here since only a subset of the outliers from the MS have been observed with ALMA. In Figure~\ref{fig:selection}, we present the SFR and stellar mass distribution of the full sample of 108 galaxies detected by $Herschel$ and having spectroscopic redshifts ($1.4<z<1.7$) from our FMOS-COSMOS program.

\subsection{Stellar mass}

Stellar masses are determined by fitting the spectral energy distribution (SED) using Hyperz \citep{Bolzonella2000} and stellar population synthesis models \citep{Bruzual2003}, based on a Chabrier IMF, at the FMOS spectroscopic redshift. Our motivation to recompute the stellar masses ourselves, for the initial selection, was to have consistency for both starbursts and MS galaxies as done in \citet{Puglisi2017}. This ensures that the enhancement in SFR for the starbursts, at a given stellar mass, are not due to difference methods applied to each galaxy type. We implement constant star formation histories while being aware that the stellar masses may be systematically lower than if assuming an exponentially-declining model with a recent burst of star formation. Photometric data covers the full spectrum from the NUV to the IR with the latter being especially important for stellar mass estimates of our highly obscured starbursts. The COSMOS field has deep NIR imaging from UltraVISTA \citep[YJHK$_{S}$;][]{McCracken2012}, Subaru Hyper Suprime-Cam (grizy; \citealt{Tanaka2017}) and $Spitzer$/IRAC \citep[3.6, 4.5, 5.8 and 8.0$\mu$m;][]{Sanders2007,Capak2017}. $Spitzer$ MIPS 24$\micron$ priors are used for de-blending Herschel (or 70$\micron$ {\it Spitzer}) sources required for accurate estimate of SFR. The stellar masses are in very good agreement with those given in \citet{Laigle2016} while considering differences between the spectroscopic and photometric redshifts. 

\begin{figure*}
\centerline{
\includegraphics[width=8cm]{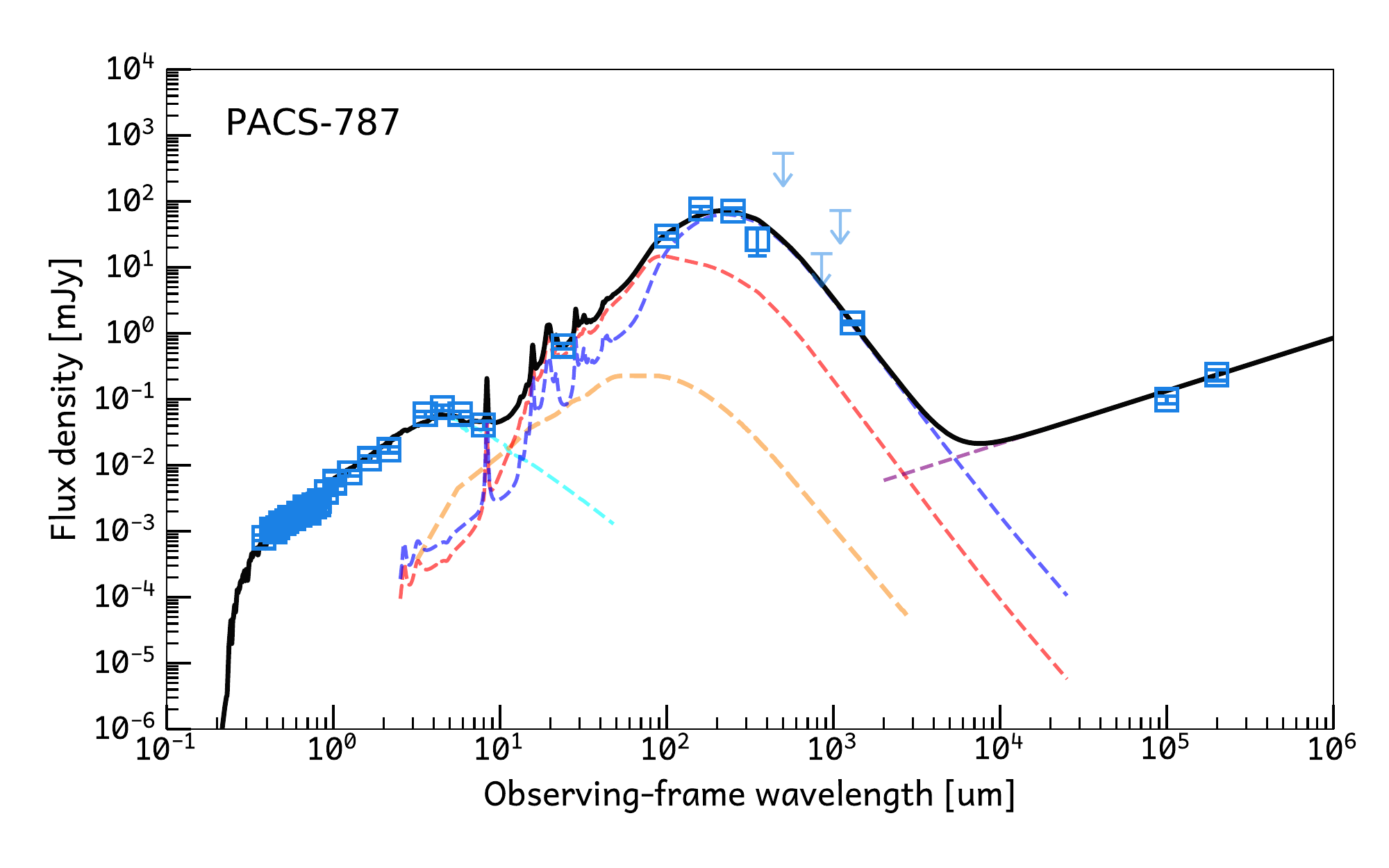}
\includegraphics[width=8cm]{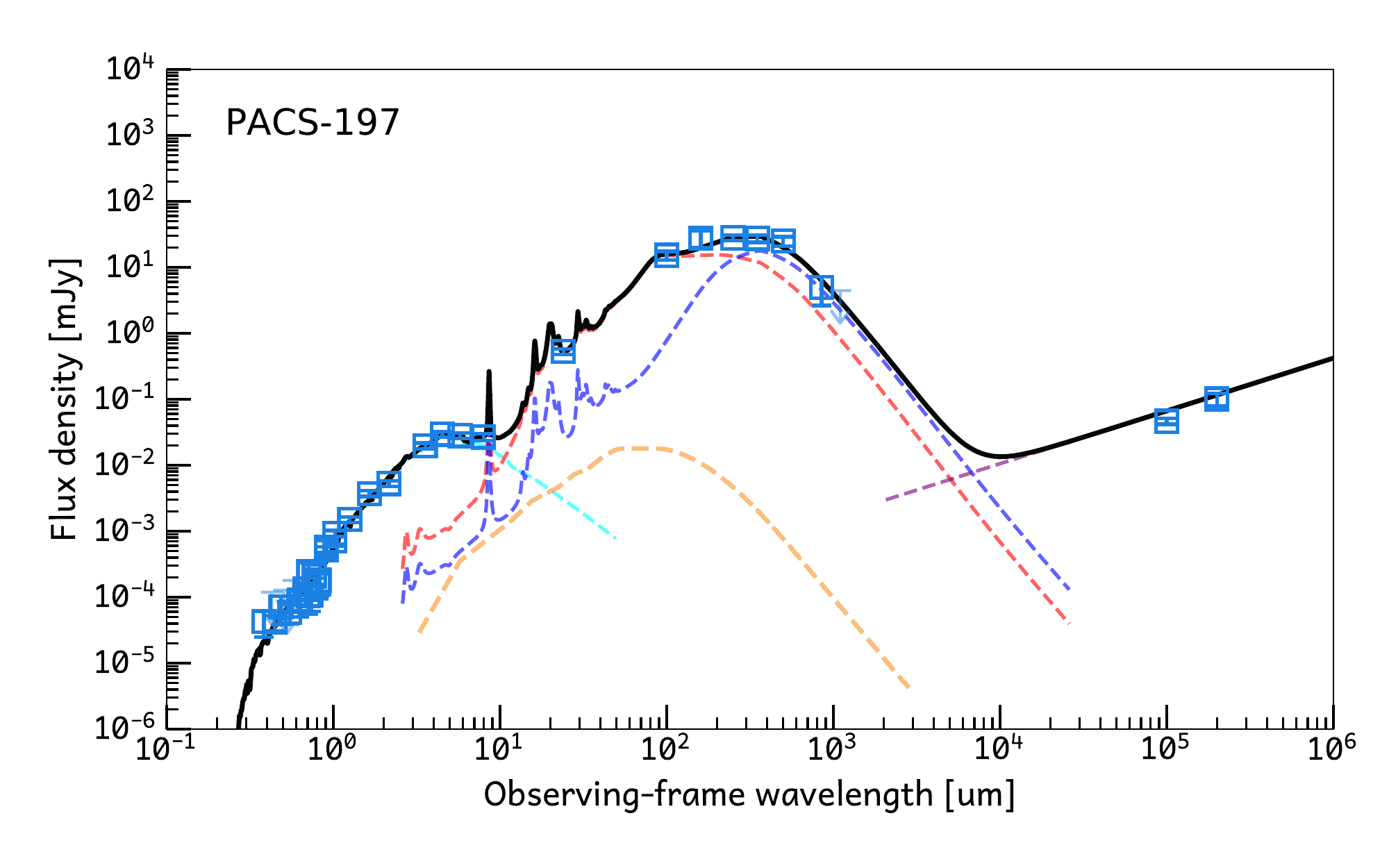}
}
\centerline{
\includegraphics[width=8cm]{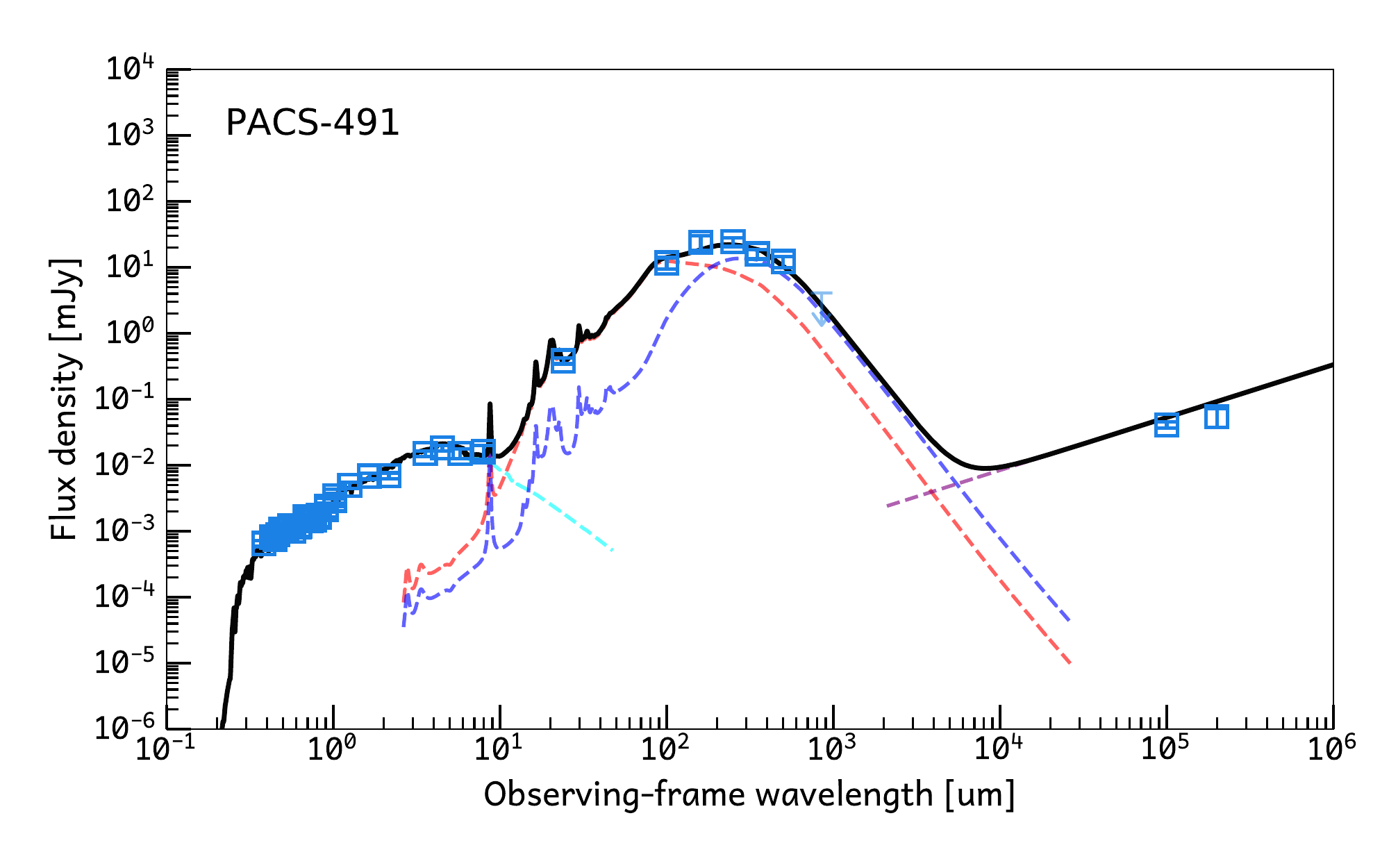}
\includegraphics[width=8cm]{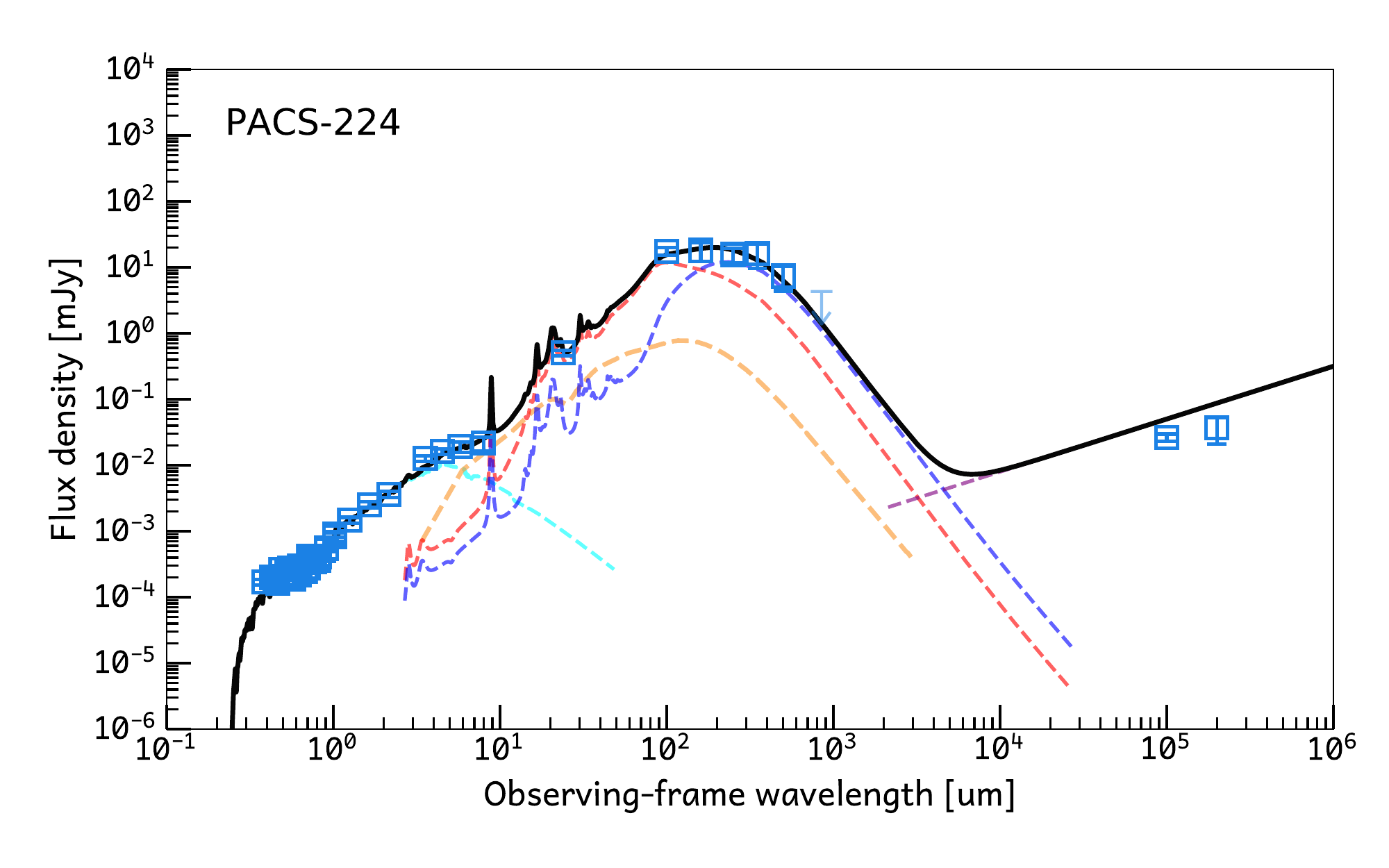}
}
\centerline{
\includegraphics[width=8cm]{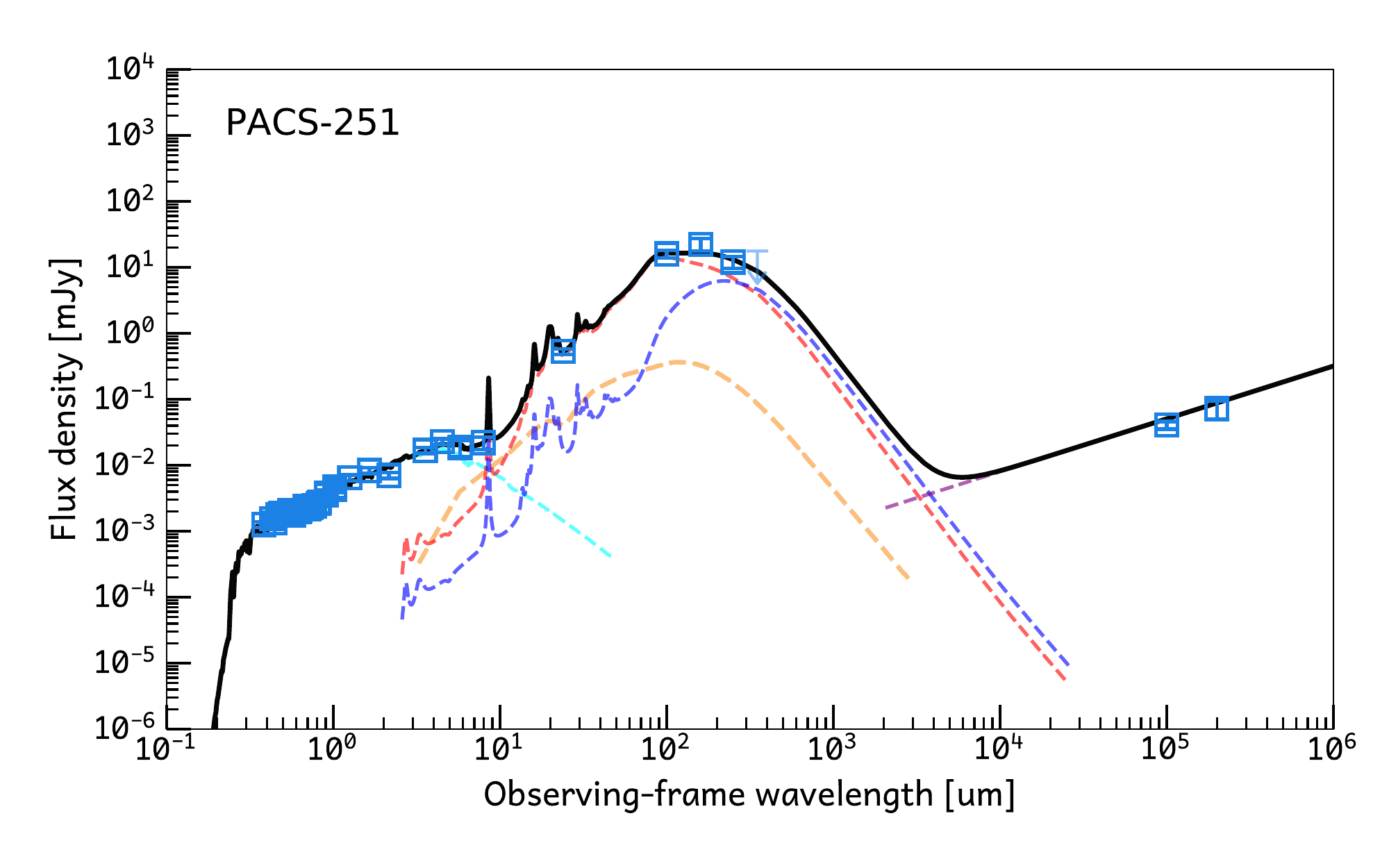}
\includegraphics[width=8cm]{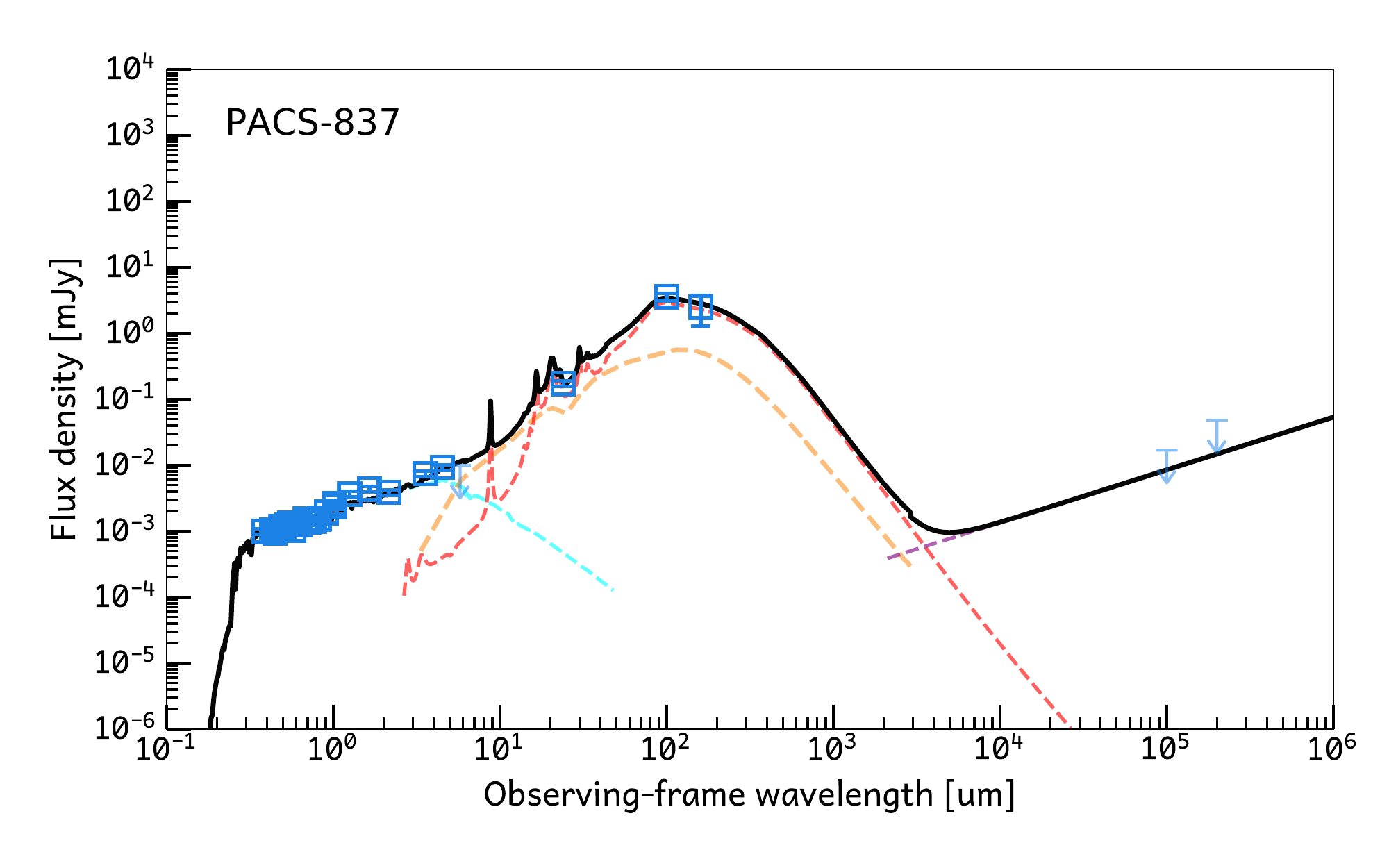}
}
\caption{SEDs of our Cycle 3 sample from IR to radio wavelengths (photometric data is shown in blue) with best-fit model shown by the black curve. The model of the dust emission is given for both a warm (red) and cool (purple) component that contributes to the mid- to far-infrared emission. The unobscured stellar emission is shown by the cyan curve. An AGN contribution (yellow) is negligible for all cases.}
\label{fig:seds}
\end{figure*}

\subsection{SFR}

For all $Herschel$-detected starbursts with spectroscopic redshifts, SFRs are determined from the total infrared (TIR) luminosity L$_{TIR}$ and the calibration of \citet{Kennicutt2012} to fully account for the obscured component that dominates the emission from starbursts \citep[$\sim90\%$;][]{Rodighiero2011}. We use the long wavelength photometry from Spitzer (24 $\mu$m; \citealt{Sanders2007}), Herschel PACS (100 and 160 $\mu$m; \citealt{Lutz2011}) and SPIRE (250, 350 and 500 $\mu$m) bands, available over the COSMOS field. In Figure~\ref{fig:seds}, we show the SEDs of our starbursts (observed with ALMA in Cycle 3). The photometric data are fit with a dust model \citep{Draine2007} with the integral from 8  - 1000 $\mu$m providing our measure of L$_{TIR}$. 

A galaxy is defined to be a starburst if the SFR exceeds, at their respective stellar mass, the mean SFR of MS galaxies by a factor $\gtrsim4$ (Figure~\ref{fig:selection}). This is effectively a selection in sSFR and not a cut on SFR alone. Our starburst sample, observed with ALMA, has SFRs ranging from $\sim$100 to 1000 M$_{\odot}$ yr$^{-1}$ (Table~\ref{tab:sample}). As a confirmation on the accuracy of the SFRs, the radio emission, as detected by the VLA at 1.4 \citep{Schinnerer2007} and 5 GHz \citep{Smolcic2017}, is in very good agreement with the power law synchrotron component as shown in each panel of Figure~\ref{fig:seds} that has a normalization only set by the infrared - radio relation. In Figure~\ref{fig:selection}, the distribution of star-forming galaxies in the $SFR-M_{stellar}$ plane illustrates the robust selection of outliers (i.e., starbursts) that fall above the star-forming MS as indicated by the best-fit linear relation (Equation~\ref{eq:MS}) to 5266 sBzK selected galaxies \citep{Daddi2004,Puglisi2017} in the COSMOS field with $1.4 < z_{phot} < 1.7$.  

\begin{equation}
\label{eq:MS}
log~SFR=0.91\pm0.09\times log~M_{stellar}-7.70\pm0.10
\end{equation}  

\noindent By selecting galaxies based on their sSFR \citep{Rodighiero2011}, we avoid the inclusion of star-forming galaxies at the massive end of the MS that would be selected if employing a method (e.g., submm flux alone) primarily sensitive to SFR. This requires an accurate determination of the star-forming MS that has been achieved with our FMOS-COSMOS sample \citep{Kashino2013}. For this purpose, we rely on accurate measurements of stellar mass and SFRs provided by the COSMOS multi-wavelength effort, particularly the deep IR imaging with UltraVISTA and $Herschel$ photometry as described above. This method allows to select objects offset ($\gtrsim4\times$ above) from the MS but less extreme than ULIRGs at low redshift. 

\subsection{Cycle 1 ALMA sample}

The five $Herschel$-detected starbursts observed in Cycle 1 with ALMA, as fully described in \citet{Silverman2015a}, were selected to have a boost in SFR greater than 4$\times$ above the MS at their respective stellar mass while the Cycle 3 observations, presented below, are selected to be further elevated from the MS. The Cycle 1 sample has spectroscopic redshifts within $1.44<z_{spec}<1.66$, stellar masses between 10$^{10}$ and 10$^{11}$ M$_{\odot}$, and SFRs greater than 300 M$_{\odot}$ yr$^{-1}$ with the exception of one of them (PACS-325) at a slightly lower SFR (Table~\ref{tab:sample}).

Total CO(2-1) emission was detected for five galaxies with a flux density for each ranging between 0.37 and  2.7 mJy and at a significance greater than 4.7$\sigma$. Having been observed at higher spatial resolution (beam size of $1.3\arcsec\times1.0\arcsec$), two sources (PACS-819 and PACS-830) show signs of extended emission while the rest are unresolved. The size and velocity measurements for these two enables us to estimate their gas mass, independent of CO luminosity, through dynamical arguments thus making an estimate of the conversion factor $\alpha_{CO}$, the ratio of molecular gas mass to CO(1 - 0) luminosity. Further details on the scientific results are given in \citet{Silverman2015a}. For the remainder of this investigation, we extend the analysis to 12 starbursts with 11 based on ALMA observations and a single observation of PACS-164 \citep{Silverman2015a} with the NOEMA interferometer. We chose not to include the CO(3 - 2) observation of PACS-282 since the galaxy is at a higher redshift ($z=2.187$) than the rest of the sample. 

\subsection{Comparison sample}
\label{sec:compsample}

\begin{figure}
\epsscale{1.2}
\plotone{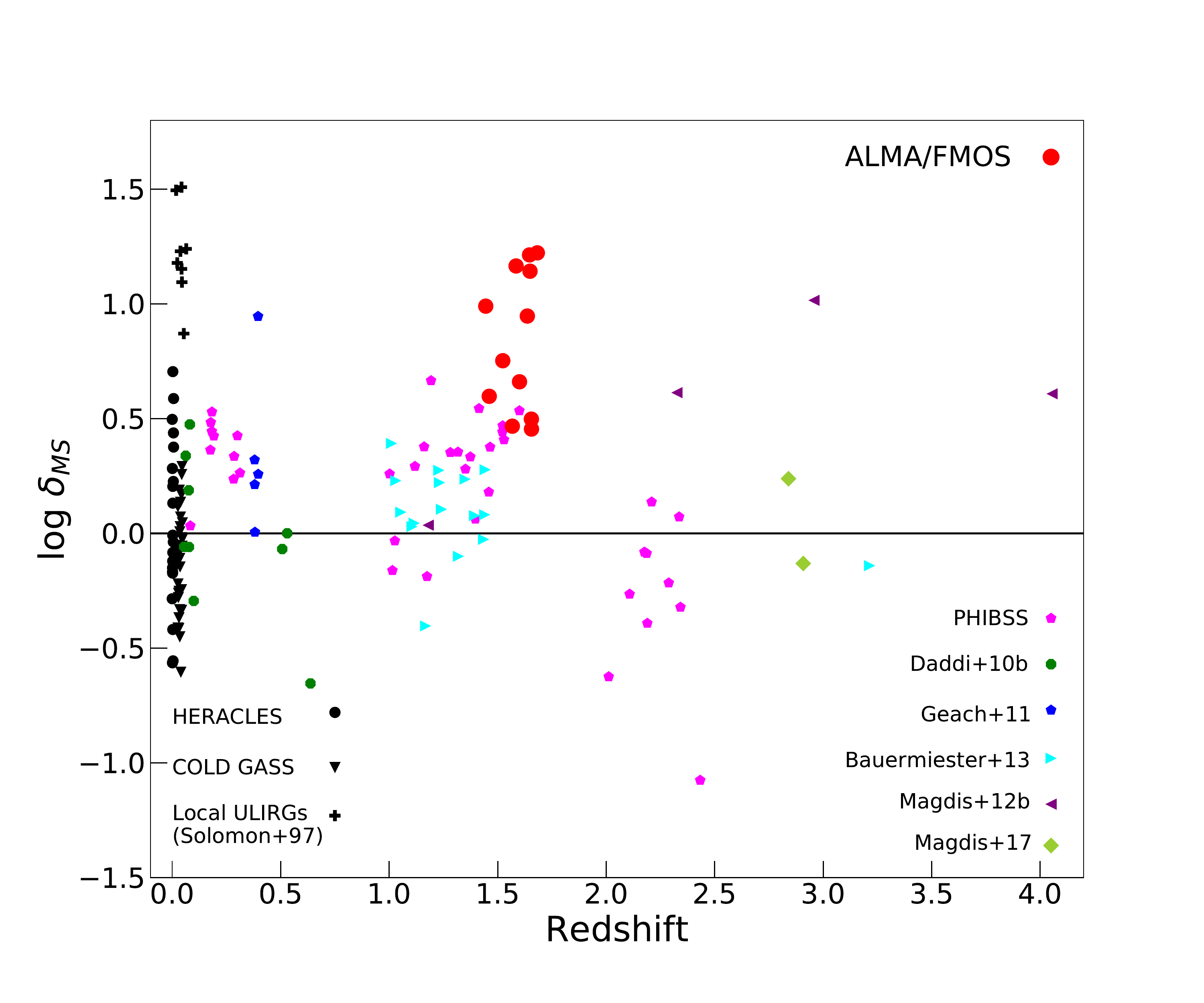}
\caption{Compilation of samples from the literature used in this analysis with their redshifts and elevation above the MS as parameterized by \hbox{$\delta_{\rm MS}$ = sSFR/$<sSFR_{MS}>$} and described in Section~\ref{text:results}. Our ALMA sample of 12 starbursts is shown by the larger red circles. Further details are provided in the text including a more extensive set of references to these data sets.}
\label{fig:comparison}
\end{figure}

For subsequent comparative analyses, a reference sample is utilized as compiled by \citet{Sargent2014} that includes 131 `typical' star-forming galaxies at $z\lesssim4$ with measurements of their CO luminosity (Figure~\ref{fig:comparison}). Here, we list and give credit to the teams responsible for these data sets. In the local Universe, the HERACLES survey \citep[e.g.,][]{Leroy2013} provides CO(2 - 1) observations of galaxies in the THINGS survey \citep{Walter2008} with the IRAM 30m single-dish telescope. COLD GASS \citep{Saintonge2011} provides CO(1 - 0) fluxes for late-type galaxies with $0.025<z<0.05$ from IRAM 30m. Higher redshift star-forming MS galaxies have been reported in many studies \citep[e.g.,][]{Geach2011,Daddi2010a,Daddi2010b,Magdis2012b,Tacconi2013,Magdis2017}. 

The starbursts included here are a restricted sample of nine local ULIRGs with two independent measurements of $\alpha_{CO}$, a dynamical assessment \citep{DownesSolomon1998} and that based on radiative transfer modeling \citep{Papadopoulos2012}. This local starburst sample has boosts above the average star-forming population of factors between $\sim10-100$ \citep{Rujopakarn2011}. In addition, three high redshift starbursts are represented: GN20 ($z=4.05$; \citealt{Tan2014}, SMMJ2135-0102 ($z=2.325$; \citealt{Swinbank2011}) and HERMES J105751.1+573027 ($z=2.957$; \citealt{Riechers2011}). We refer the reader to \citet{Sargent2014} and \citet{Tacconi2018} for a complete description of the individual samples. In Figure~\ref{fig:comparison}, we also include our ALMA/FMOS starburst sample at $z\sim1.6$. It is evident that this study improves the statistics with respect to the number of starbursts at $z>1$ with CO detections; this is also exemplified in Figure 2 of \citet{Tacconi2018}.  

We mention that the published samples used in this analysis were chosen to well represent the molecular gas properties of star-forming galaxies over a wide redshift range for comparison with our high-z starbursts. There are additional samples that are available \citep[e.g.,][]{Combes2013,Bothwell2013} but not included here since our aim was not to present a comprehensive assessment of the field.  For this, we refer the reader to the recent work by \citet{Tacconi2018}.

\section{ALMA observations and data analysis}

Seven galaxies were observed in Cycle 3 (Project 2015.1.00861.S) with 38 - 39 12m antennas on 4, 11, and 13 of March 2016. Slightly different configurations for each science block yielded an angular resolution between 1.7-1.8$\arcsec$, slightly higher than the request of 3$\arcsec$. We selected spectral windows to detect $^{12}$C$^{16}$O(2 - 1; $\nu_{rest}$=230.538 GHz) redshifted into Band 3. Four base bands ($\Delta\nu=1.875$ GHz each) were configured to detect CO emission for multiple targets (2 - 3) at different redshifts within a single science block and provide a measure of the continuum as an additional science product. Standard targets were used for calibration (e.g., flux, phase, bandpass) including J0854+2006, J0948+0022, Callisto, and J1058+0133. 

Based on ALMA data from another program (PI. E. Daddi; Project 2015.1.00260), one target (PACS-472; RA=10:00:08.95, DEC=02:40:10.8) has a clear CO emission-line detection in Band 6 (private communication) that places the object at a higher redshift than that based on our FMOS spectra. This misidentification explains a lack of a CO detection at the expected observed frequency thus we remove it from the sample. In Table~\ref{tab:sample}, we list the targets and give their physical characteristics (e.g., redshift, stellar mass, SFR). 

Analysis of the interferometric data set is carried out using the standard analysis pipeline available with CASA Version 4.6 \citep{McMullin2007}. We first generate an image of the CO emission using the task `immoments' with a channel (i.e., velocity) width encompassing the full line profile as given in Table~\ref{tab:co}. The emission is then modeled with an elliptical Gaussian using the CASA tool `imfit' that returns a source centroid, deconvolved source size, and integrated flux. As a precaution, we confirm these measurements by fitting the emission in the $uv$-plane and through an independent effort using GILDAS available in the MAPPING package.

\begin{figure*}

\gridline{\fig{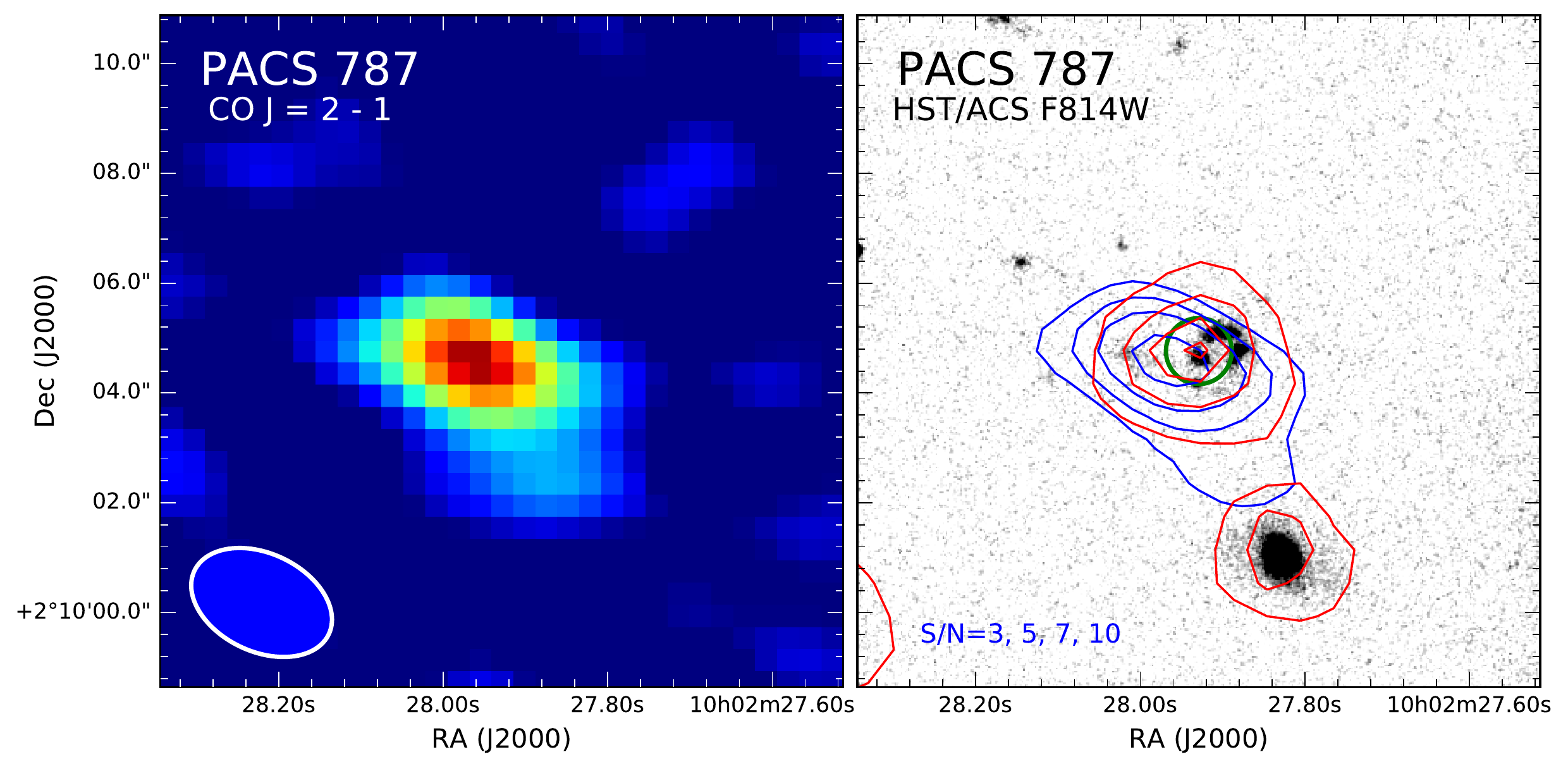}{0.5\textwidth}{}
          \fig{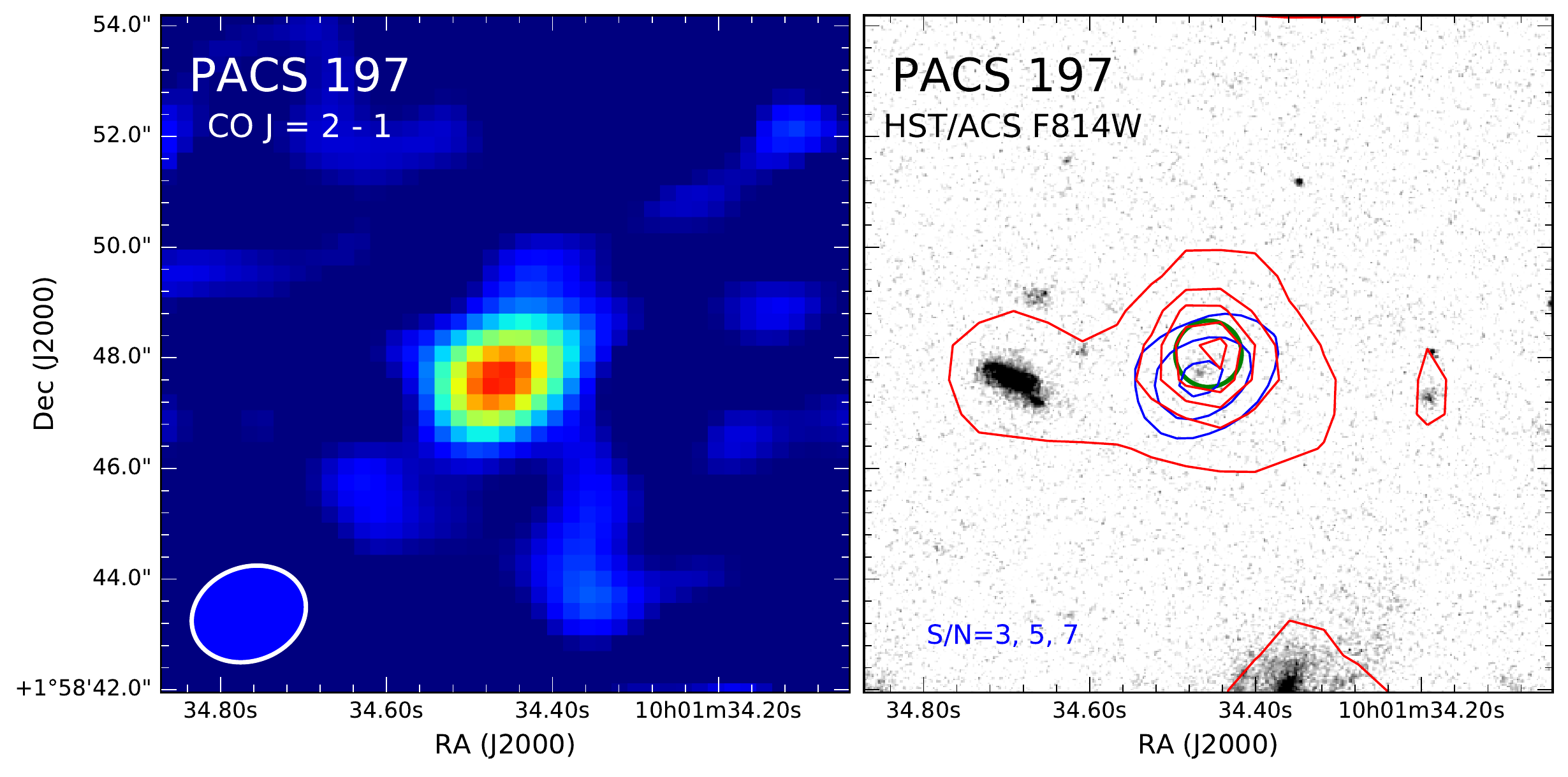}{0.5\textwidth}{}       
          }
\gridline{\fig{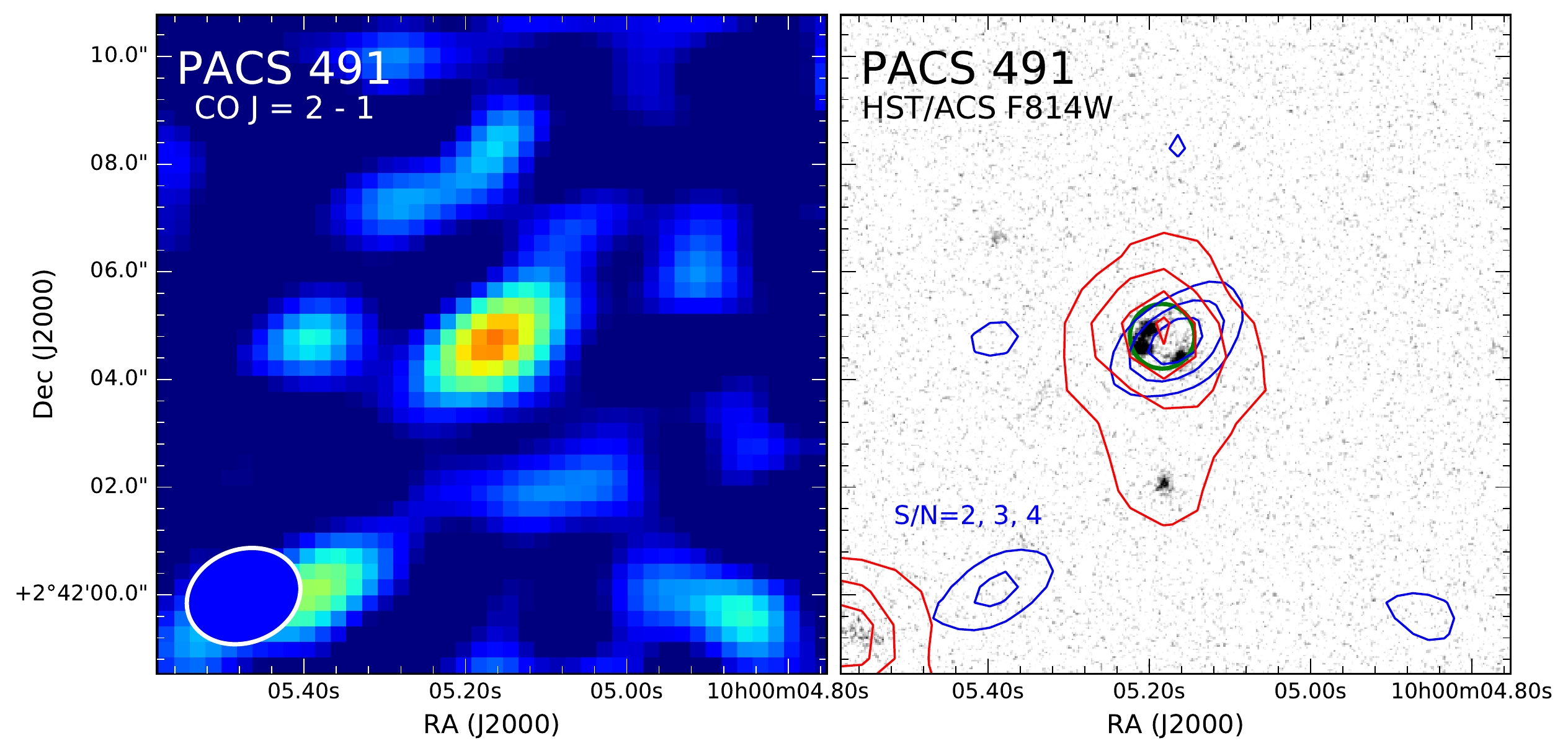}{0.505\textwidth}{}
             \fig{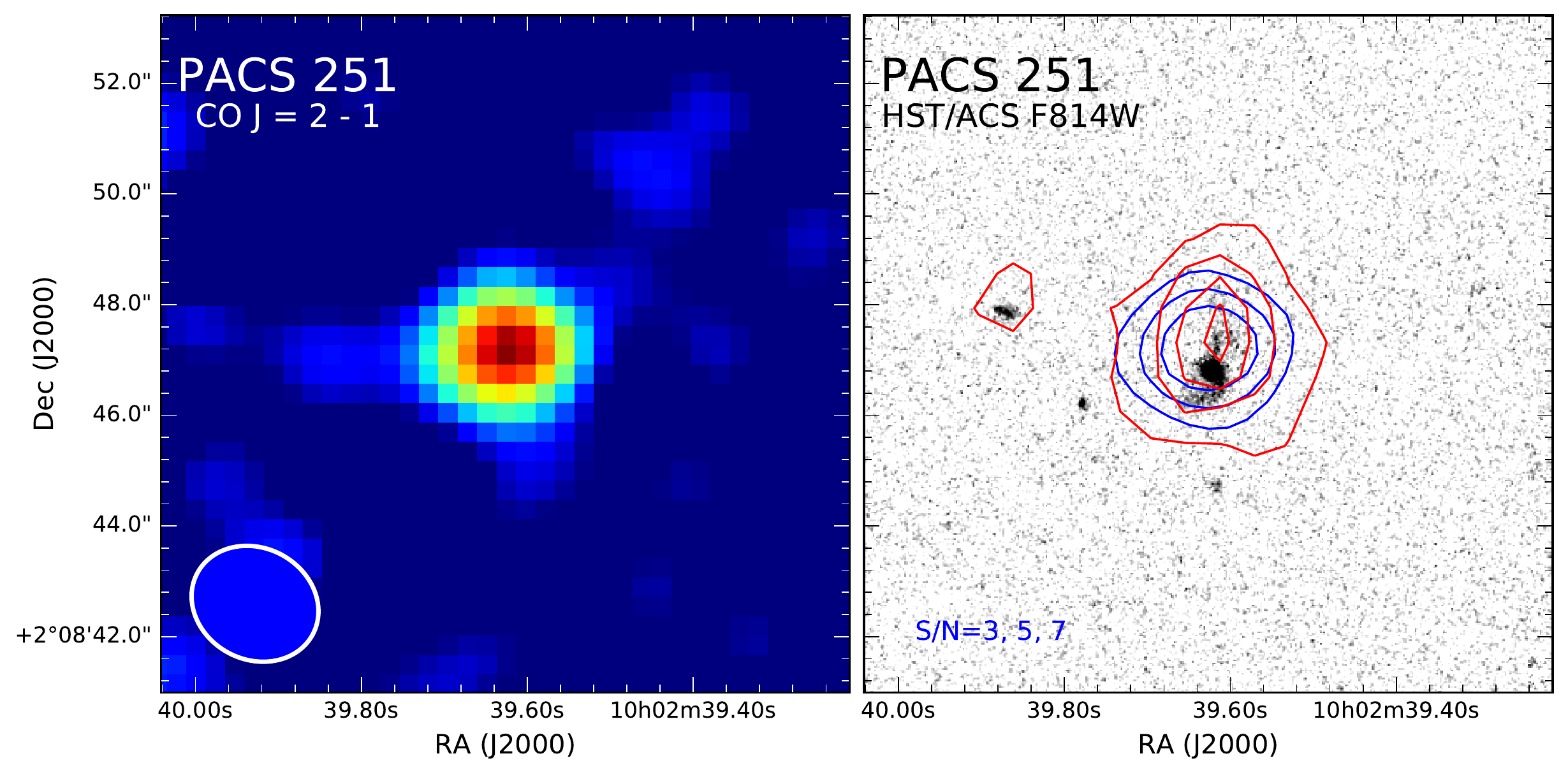}{0.5\textwidth}{}
}
\gridline{
\fig{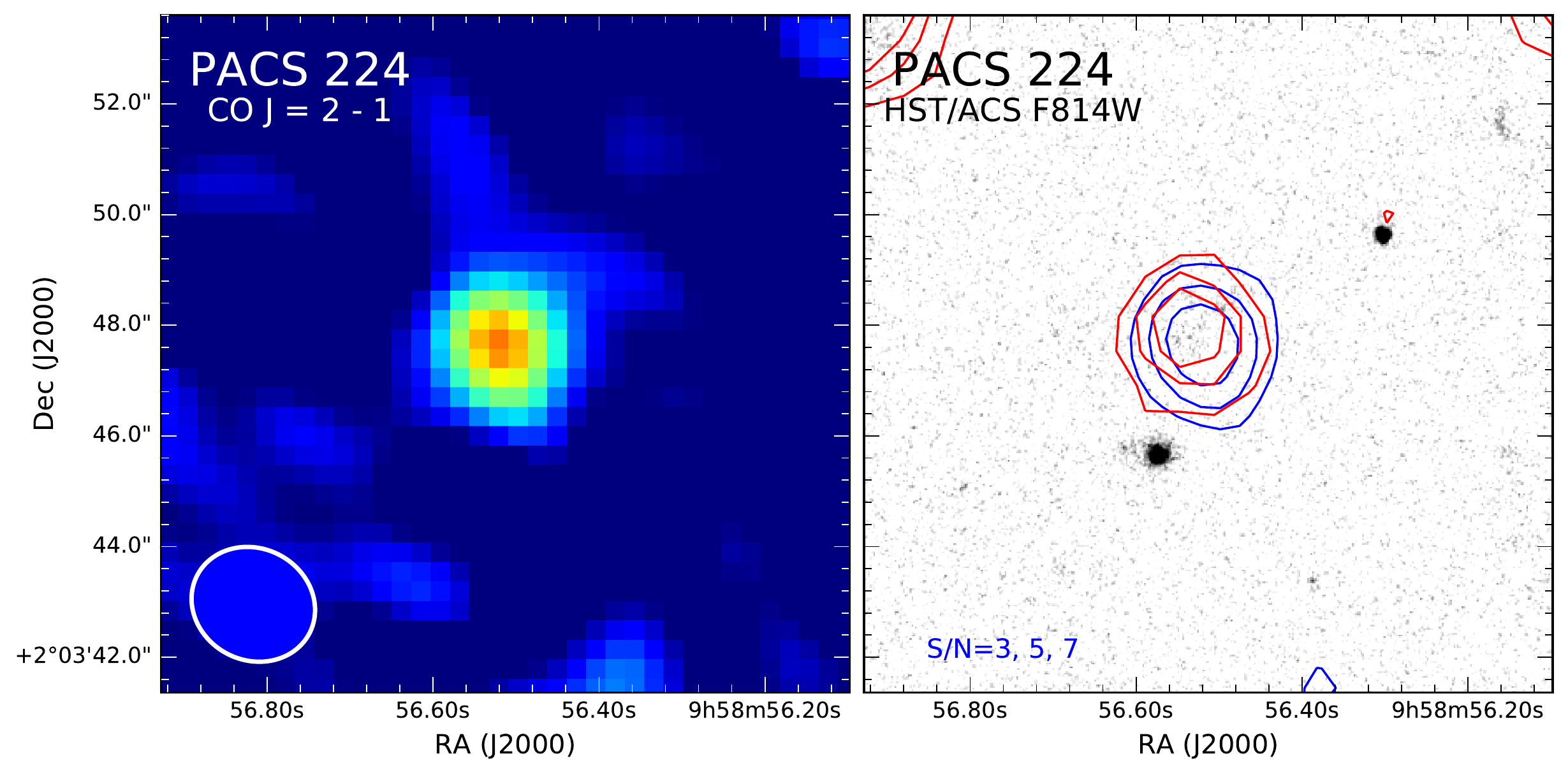}{0.50\textwidth}{}
\fig{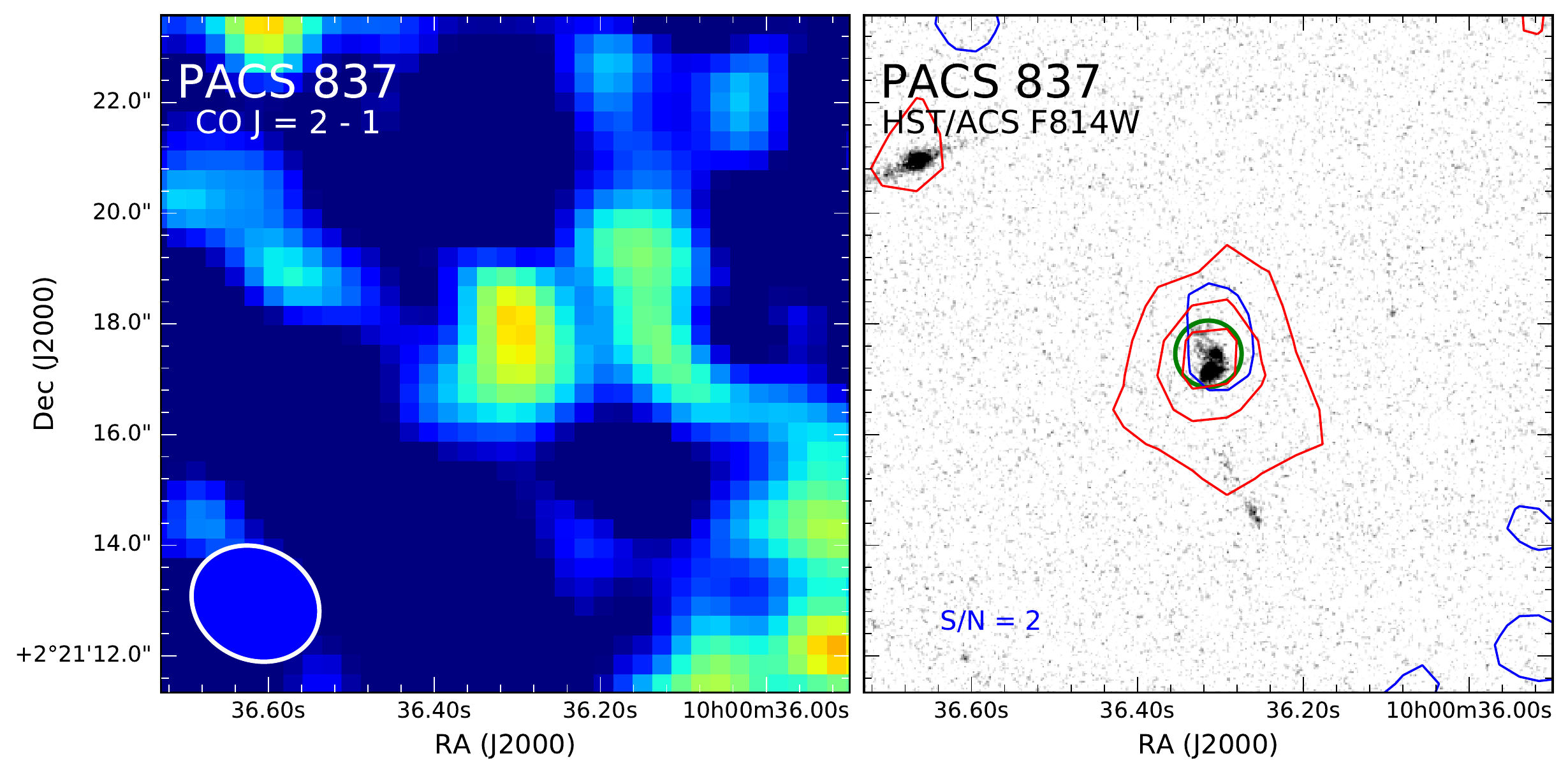}{0.5\textwidth}{}
  	  }	  

\caption{CO(2 - 1) maps of six high-z starbursts observed by ALMA in Cycle 3 (color panels on left). The shape of the ALMA beam is displayed in each panel. The minimum flux level is set at $0.5\times\sigma_{rms}$ (Table~\ref{tab:co}). $Right$ panels: Greyscale images of each starburst, from the COSMOS HST/ACS F814W mosaics \citep{Koekemoer2007}. CO emission is indicated with blue contours in steps of S/N as shown in each panel. Red contours indicate $Spitzer$/IRAC 3.6$\mu$m detections that are typically co-spatial with the CO emission indicating a close association between gas, obscured star formation and the peak of the stellar mass distribution. The green circle in each HST panel marks the position of the FMOS fiber with a diameter of 1.2$\arcsec$.} 
\label{fig:co_images}
\end{figure*}

\section{Results}
\label{text:results}

\begin{figure}
\epsscale{1.3}
\plotone{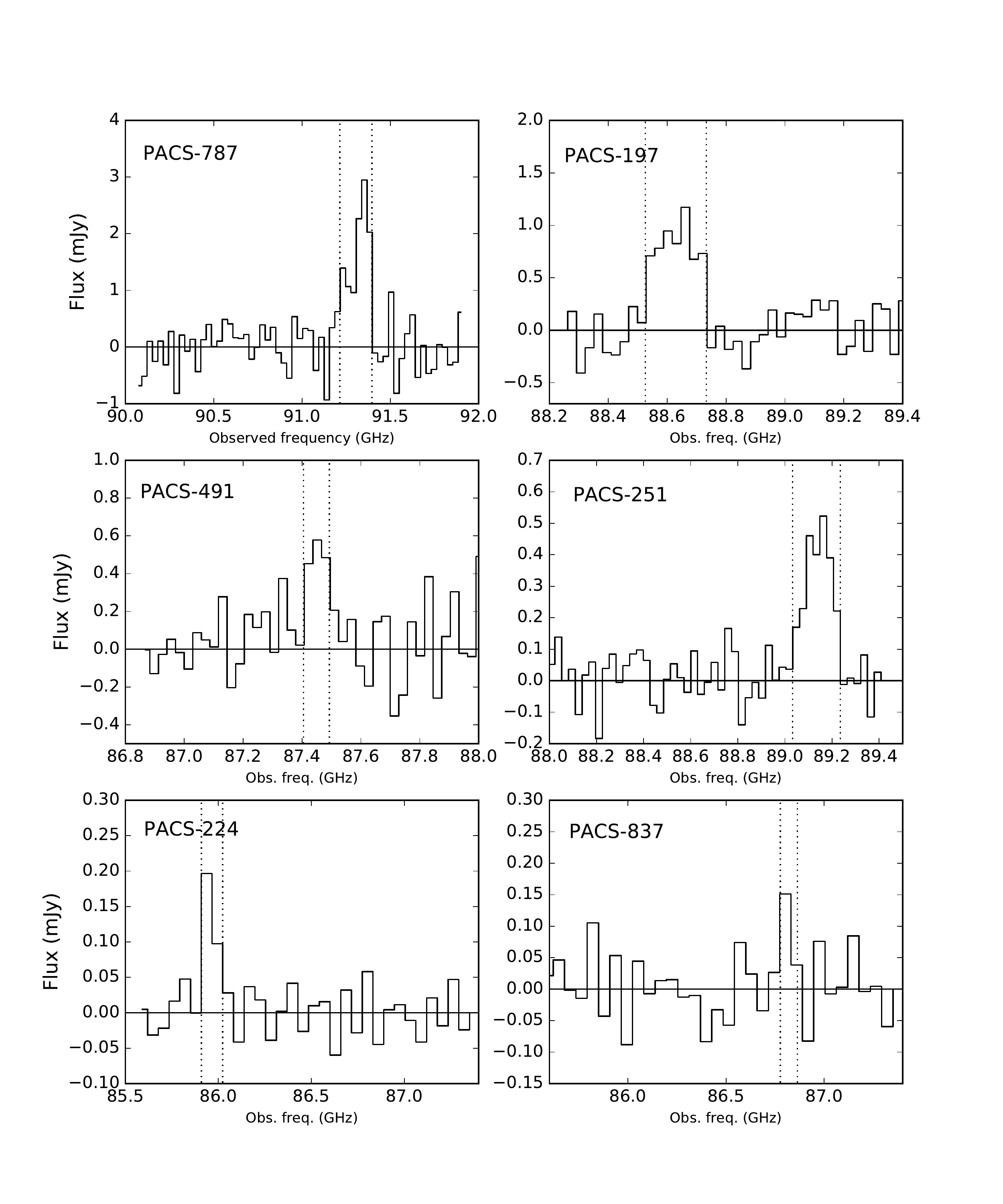}
\caption{CO(2 - 1) spectra for the Cycle 3 ALMA sample. Observed velocity channels are binned in intervals of 100 (PACS-787, 197, 491, 251) or 200 (PACS-224, 837) km s$^{-1}$. Spectra were extracted with different apertures for each source chosen to closely represent the unresolved (i.e. peak) CO emission. The vertical dotted lines indicate the velocity interval over which the total CO luminosity is measured as given in Table~\ref{tab:co}. The horizontal line marks the zero level.}
\label{fig:co_profiles}
\end{figure}

We detect CO at $S/N>7$ in five out of six galaxies observed in Cycle 3 (Figure~\ref{fig:co_images}; Table~\ref{tab:co}). The line intensities (I$_{\rm CO}$) span a range of 0.26 to 1.64 Jy km s$^{-1}$ with the brightest source in CO emission attributed to our most extreme outlier from the MS (PACS-787; SFR = 991 M$_{\odot}$ yr$^{-1}$). In Figure~\ref{fig:co_profiles}, the velocity profiles of the CO detections are shown in bins of either 100 or 200 km s$^{-1}$. The majority have significant detections across multiple velocity bins. Overall, the strength of the CO emission is indicative of large amounts of molecular gas out of which deeply-embedded stars are forming (see Section~\ref{text:gasfractions}). 

All five sources with CO detections are essentially unresolved with the beam sizes given in the Table~\ref{tab:co}. However, we were able to measure a size for PACS-787 of $1.96\pm0.54\arcsec~(16.6~{\rm kpc; FWHM-major~axis}) \times1.02\pm0.39\arcsec~(8.6~{\rm kpc; FWHM - minor~axis})$ based on a elliptical Gaussian fit. We now know that the CO(5 - 4) emission from PACS-787, based on higher resolution imaging with ALMA in Band 6, is nearly equally distributed between two galaxies undergoing a major merger, each with compact ($r_{1/2}\sim1$ kpc) disks and having a separation of 8.6 kpc (Silverman et al. 2018, in press).    

PACS-837 does not have significant emission in either CO or the continuum, although there is a tentative CO detection (0.42$\pm$0.20 mJy) of 2.1$\sigma$ significance at the expected spatial location and redshift ($z_{CO}=1.6569$). For these reasons, we include a panel for this object in Figures~\ref{fig:co_images} and~\ref{fig:co_profiles}. In subsequent analyses, we have derived a 3$\sigma$ upper limit that places important constraints on our characterization of the CO properties of the sample since it should have been detected at a higher significance if having CO properties similar to starburst galaxies given their L$_{\rm TIR}$. 

Below, we present derived quantities including CO luminosity, gas mass, and gas depletion time as a function of SFR and stellar mass. To aid in the interpretation of our results, we also present these measurements normalized to the expected value for MS galaxies at their respective redshift and stellar mass using average relations available from the literature. In particular, the expected value of the CO luminosity for MS galaxies ($<L^{\prime}_{CO, MS}>$) is based on Equation 1 of \citet{Sargent2014} that depends on the $L_{TIR}$ of each galaxy. A similar relation is given in Equation 4 of \citet{Sargent2014} for the mean molecular gas mass of MS galaxies as a function of SFR. We then compare these normalized quantities as a function of their sSFR relative to the mean sSFR ($<sSFR_{MS}>$) of MS galaxies at an equivalent redshift and stellar mass such that \hbox{$\delta_{\rm MS}$ = sSFR/$<sSFR_{MS}>$}. This normalization scheme applies to all figures that present the measurements in terms of $\delta_{\rm MS}$ including Figure~\ref{fig:comparison}. For our ALMA starbursts at $z\sim1.6$, we use the MS relation as given in Equation~\ref{eq:MS}. To show the effect of using a different parameterization of the MS at $z\sim1.6$, we also show results using the definition of the MS from \citet{Speagle2014} in the following subsection only. For the comparsion sample, we use the parameterization of $sSFR~(M_{stellar}, z)$ as given in Equation A1 of \citet{Sargent2014}.

\begin{figure}
\epsscale{2.5}
\plottwo{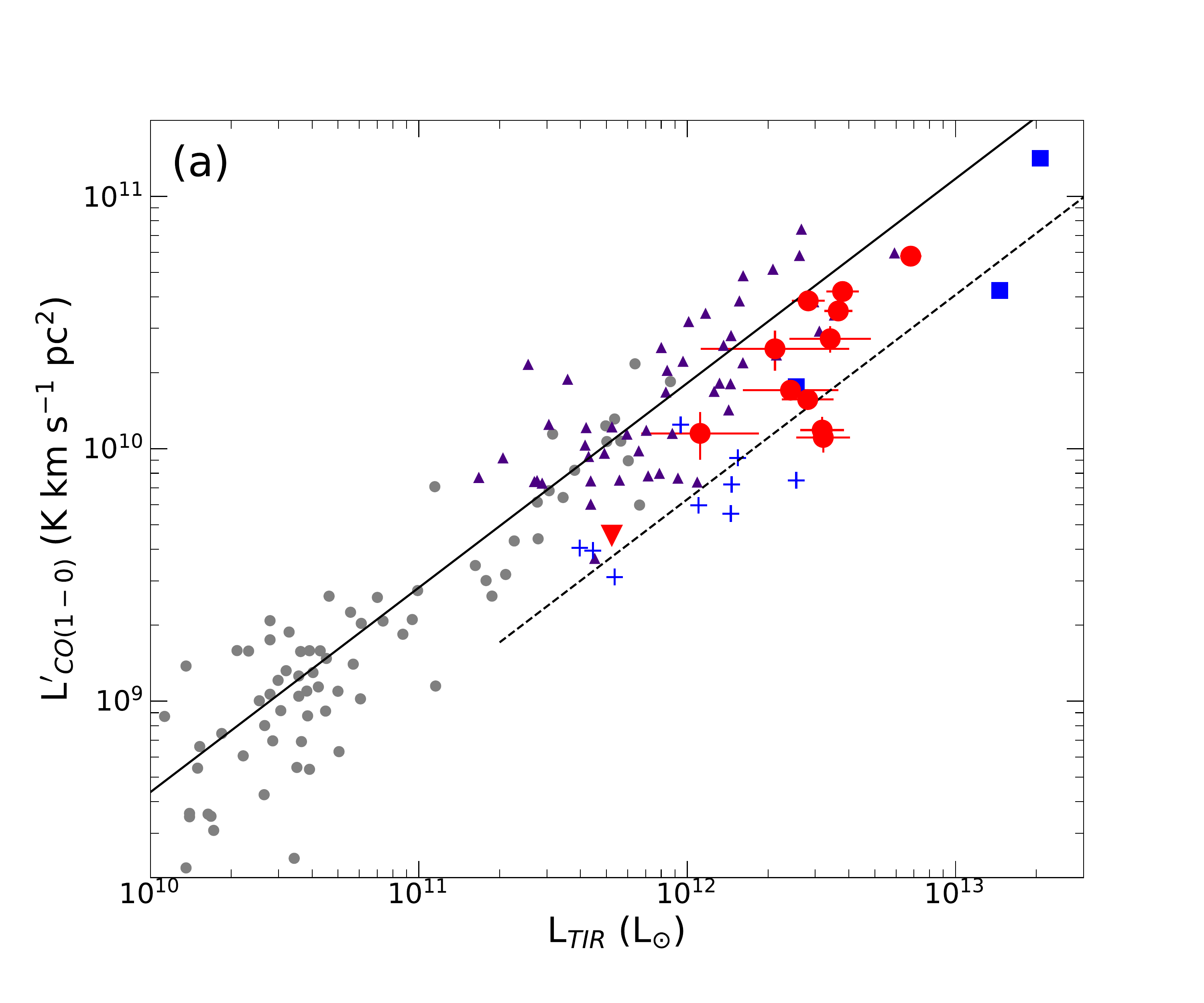}{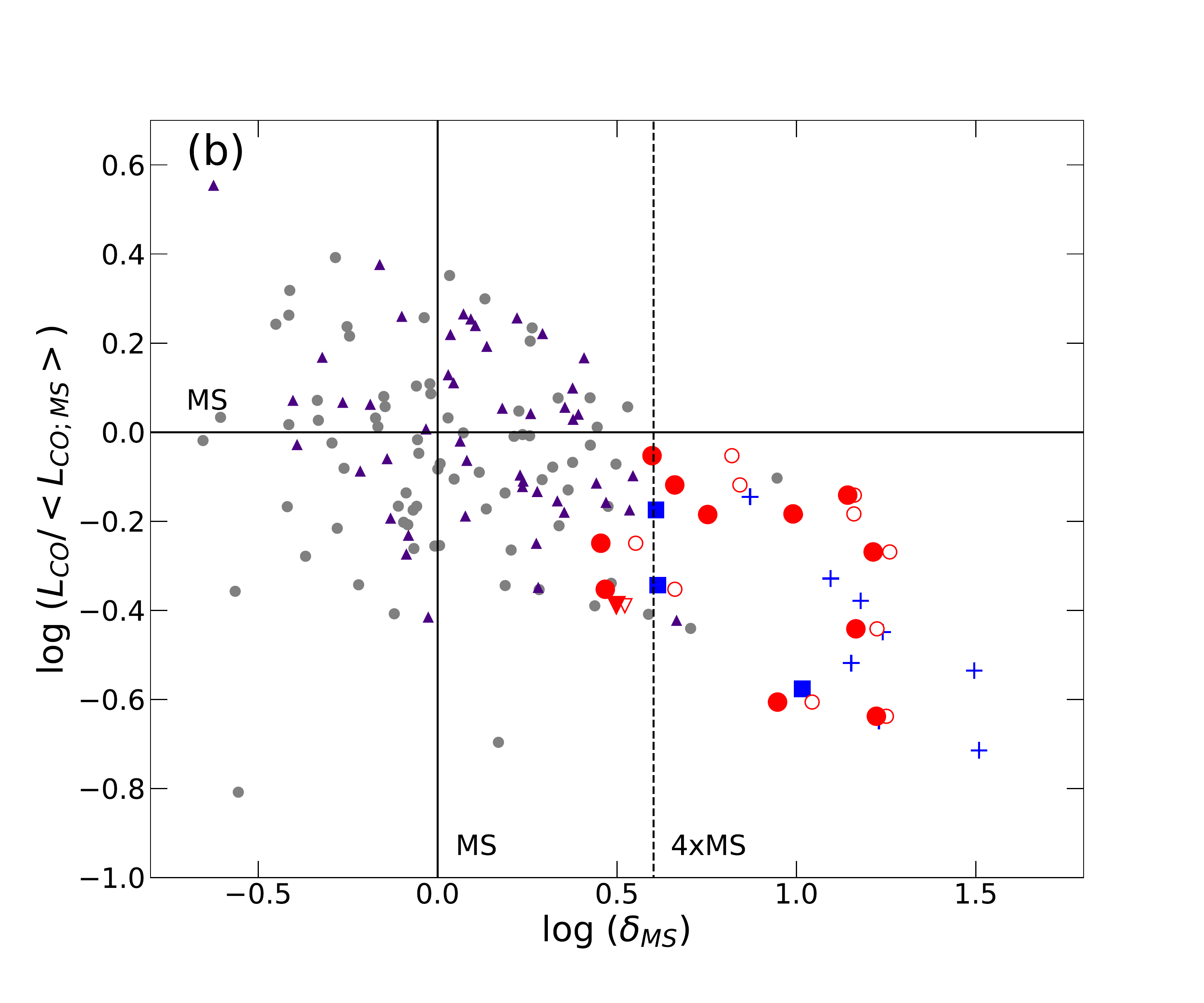}
\caption{(a) CO(1 - 0) luminosity ($L^{\prime}_{CO}$) as a function of total infrared luminosity ($L_{TIR}$) and (b) CO luminosity and sSFR both normalised to the mean value of typical star-forming MS galaxies as described in the text. Red (filled) data points mark our twelve high-z starbursts. The inverted triangle represents the 3$\sigma$ upper limits for PACS-837. The red open symbols are equivalent to the red filled symbols with the exception that the value of $\delta_{\rm MS}$ is in reference to the star-forming MS of \citet{Speagle2014}. From the compilation of \citet{Sargent2014} as described in Section~\ref{sec:compsample}, the small grey circles (purple triangles) represent star-forming MS galaxies at $z<1$ ($z>1$) while blue symbols show the local ULIRGs (crosses) and high-z starbursts (filled blue squares). In panel (a), the best-fit relation to MS galaxies is shown by the solid line with a similar relation shifted $3\times$ lower to indicate the location of local ULIRGs \citep{Sargent2014}.}
\label{fig:tir_co}
\end{figure}

\subsection{CO-to-TIR luminosity relation}

We revisit the relation between the CO luminosity $L^{\prime}_{CO}$ and $L_{\rm TIR}$ for starbursts as reported by many studies \citep[e.g.,][]{Magdis2012a,Sargent2014}. For our sample, the CO line luminosity is calculated as follows and given in units of K km s$^{-1}$ pc$^2$  

\begin{equation}
L^{\prime}=3.25\times10^7\times S_{CO} \Delta v \frac{D_L^2}{(1+z)^3 \nu_{obs}^2}
\end{equation}  

\noindent where $S_{CO}$$\Delta$$v$ is the line flux in units of Jy km s$^{-1}$, $D_L$ is the luminosity distance in Mpc, and $\nu_{obs}$ is the observed line frequency in GHz. We then convert the luminosity to the value of the $J= 1 - 0$ transition using $L^{\prime}_{CO(2-1)}$/$L^{\prime}_{CO(1-0)}=0.85$ \citep{Daddi2015}. As mentioned above, $L_{\rm TIR}$ is determined from an integral of the best-fit model over 8-1000 $\mu$m, which tightly correlates with the obscured SFR \citep{Kennicutt98}. Before introducing additional uncertainties with converting to physical units, we point out that these quantities $L^{\prime}_{\rm CO}$ and $L_{\rm TIR}$ are essentially based on flux measurements and distance measures from their spectroscopic redshifts with the exception of converting between CO transitions. This allows us to establish observable trends independent of conversion factors to quantities such as gas mass and SFR that will be presented in the following section.
 
We plot $L^{\prime}_{CO}$ versus $L_{\rm TIR}$ for our starburst galaxies (Figure~\ref{fig:tir_co}$a$) and a comparison sample of galaxies as described in Section~\ref{sec:compsample}. The well-established linear relation (solid line) is seen for which all MS galaxies lie along \citep{Sargent2014}. As presented by others \citep[e.g.,][]{Solomon1997,Daddi2010b,Genzel2010}, the local ULIRGs (shown by the blue crosses) are offset from this relation with lower CO luminosities at a given $L_{\rm TIR}$ as further illustrated by a parallel relation (dashed line), a factor of $3\times$ below the MS relation. Pertaining to our high-z starbursts, the entire ensemble (large red symbols) is visibly displaced to lower $L^{\prime}_{CO}$ than expected for MS galaxies at their respective $L_{TIR}$, even though a few galaxies do fall close to (but below) the mean relation for MS galaxies. While some of our high-z starbursts lie along this parallel track (dashed line) to the MS, similar to the local ULIRGs, our sample appears to fill in the region between the MS and local starburst galaxies thus not as extreme in their difference from the MS galaxies.

In Figure~\ref{fig:tir_co}$b$, we illustrate that the decrement in CO luminosity for high-z starbursts is larger with increasing boost in SFR above the MS by plotting this quantity as a function of sSFR, with each quantity normalized to the mean value of the star-forming MS population as described above. As shown in the figure, there is a general decline in L$^{\prime}_{\rm CO}$ / $<$L$^{\prime}_{\rm CO,MS}$$>$ with increasing \hbox{$\delta_{\rm MS}$}. Since there are no clear signs of a gap in the ratio L$^{\prime}_{\rm CO}$ / $<$L$^{\prime}_{\rm CO, MS}$$>$ between the MS and starburst galaxies, we argue that this decline is continuous in these parameters. This result appears to be valid even if using the parameterization of the MS at $z\sim1.6$ from \citet{Speagle2014} that shows slightly higher boost factors. These results are likely enabled by our selection of starbursts that includes those with milder offsets from the MS as compared to the local ULIRGs, thus effectively filling the gap seen in other studies \citep{Daddi2010b,Genzel2010}. 

Based on a Kolmogorov-Smirnov test, there is only a probability of 0.07\% of randomly drawing values of L$^{\prime}_{\rm CO}$ / $<$L$^{\prime}_{\rm CO, MS}$$>$  from the distribution of normal star-forming galaxies that matches the distribution of our starbursts with $sSFR/<sSFR_{MS}> ~ > 4$. Furthermore, the difference in the mean of the L$^{\prime}_{\rm CO}$ / $<$L$^{\prime}_{\rm CO, MS}$$>$ distribution between typical galaxies and our starburst sample is significant at the $13\sigma$ level. This decline of the L$^{\prime}_{\rm CO}$/L$_{\rm TIR}$ ratio suggests that starburst galaxies are able to sustain high levels of star formation without the need for a larger gas supply, hence supporting a scenario of a higher star-formation efficiency (SFE=SFR/M$_{\rm gas}$; see below) {\it in a continuous manner from MS galaxies to the most extreme starbursts} as seen in the local ULIRGs. As an alternative explanation, if molecular clouds in starbursts are simply denser as compared to MS galaxies, the reduced CO luminosity may be attributed to the lower surface area of the clouds since CO lines are optically thick.  

\subsection{Total molecular gas mass}
\label{text:gasfractions}

We expand on the above results by converting the CO luminosity, $L^{\prime}_{CO(1-0)}$, to the total molecular gas mass ($M_{gas}$) using a scale factor ($\alpha_{CO}$) as routinely done in the literature \citep{Bolatto2013}. A single value of this factor ($\alpha_{CO}=1.3$ M$_{\odot}$ / (K km s$^{-1}$ pc$^{2}$)) is applied across our starburst sample. This estimate of $\alpha_{CO}$ is based on a dynamical assessment of the gas mass using a higher resolution observation with ALMA of CO(5-4) emission from PACS-787 (Silverman et al. 2018, in press). This value of $\alpha_{CO}$ is similar with that reported in the literature for local \citep{Sargent2014} and high-redshift starburst galaxies \citep[e.g.,][]{Hodge2015} including two other cases in our sample \citep{Silverman2015a}. We recognize that different values of $\alpha_{CO}$ have been assumed in the literature. For example, \citet{Tacconi2018} use a single value of $\alpha_{CO}=4.36$ (slightly depending on metallicity) for all MS and starburst galaxies alike (i.e., for all $\delta_{\rm MS}$ values). To assess the impact of this difference on our results, we present our analysis in all subsequent plots also with this higher value for $\alpha_{CO}=4.36$ M$_{\odot}$ / (K km s$^{-1}$ pc$^{2}$) and discuss the implications in Section~\ref{sec:discussion}. We highlight that the uncertainty on the appropriate value of $\alpha_{CO}$ is the dominant systematic error in our measure of derived properties that include the molecular gas mass.

\begin{figure}
\epsscale{2.5}
\plottwo{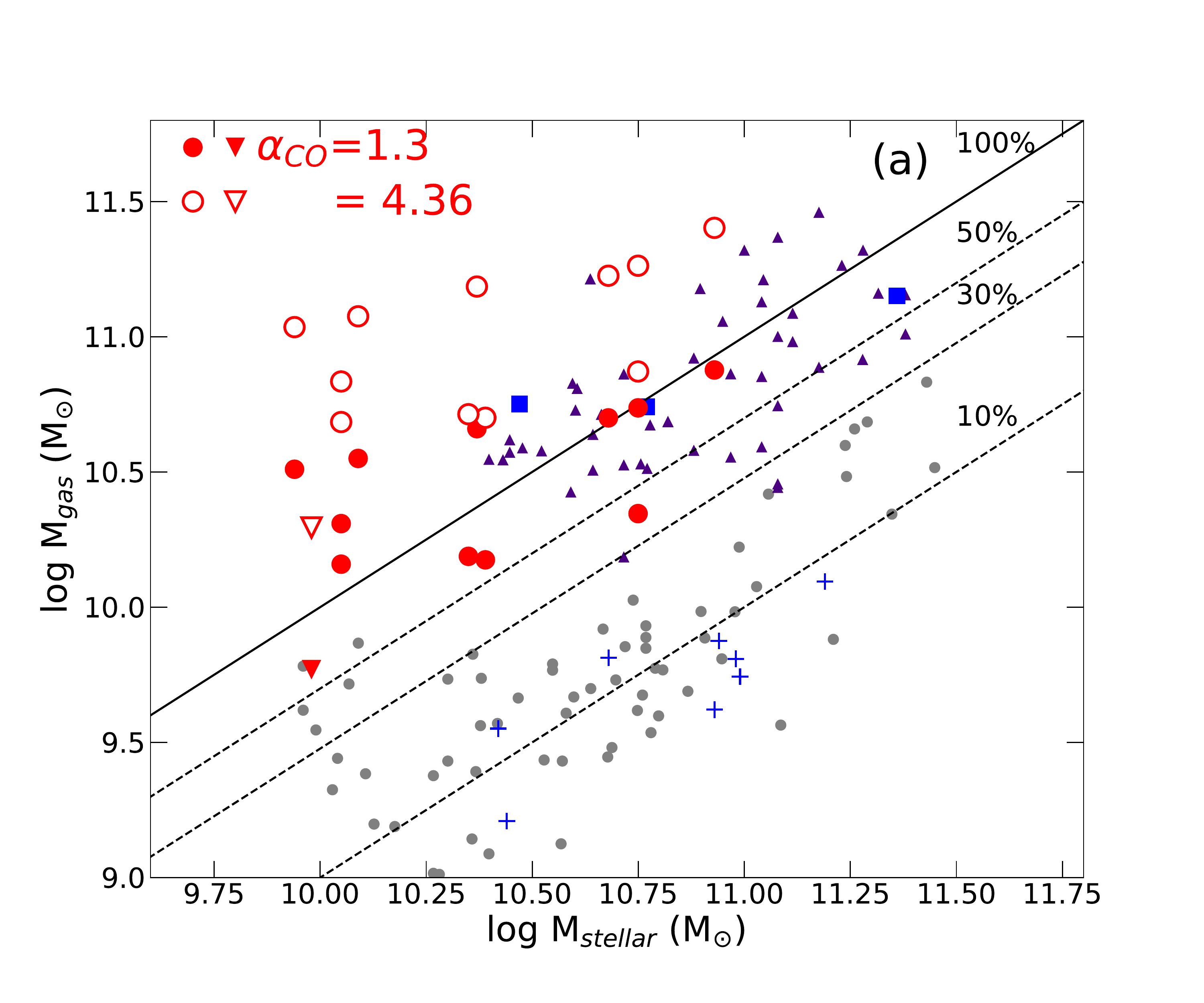}{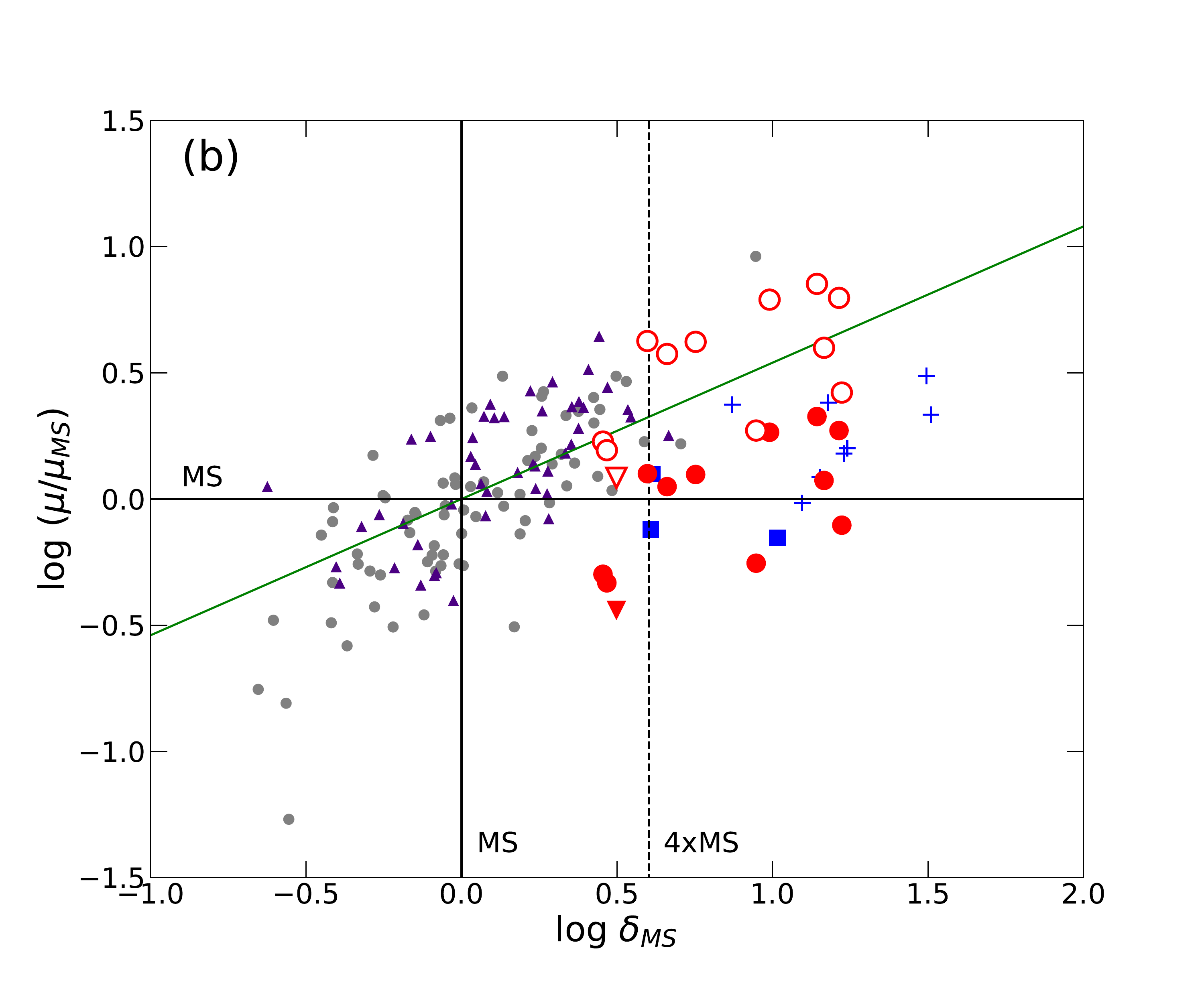}
\caption{Molecular gas mass of high-z starbursts. (a) Log of the molecular gas mass as a function of the log of the stellar mass. Our ALMA starburst sample is shown by filled (open) red symbols, assuming the value of $\alpha_{CO}$ to be 1.3 (4.36) M$_{\odot}$ / (K km s$^{-1}$ pc$^{2}$) and an inverted triangle indicates an upper limit on the gas mass. Slanted lines indicate increasing values of M$_{gas}$/M$_{stellar}$ as marked. Additional starburst galaxies from the literature (listed in Section~\ref{sec:compsample}) with $\alpha_{CO}$ measurements are included at low (blue crosses) and high (blue squares) redshift. Small grey circles represent low-redshift ($z<0.3$) MS star-forming galaxies while high-z ($z>1$) MS galaxies are marked by purple triangles. (b) Molecular gas mass versus sSFR, with both quantities normalized to the mean value of SF MS galaxies, at the equivalent stellar mass. The parameter on the abscissa, $\delta_{\rm MS}$, is the same as in Figure~\ref{fig:tir_co}b. The slanted green line is the average relation as reported in \citet{Tacconi2018}. The grey circles in the lower panel represent all SF MS galaxies at $z<1$, a less restrictive sample than in the top panel.}
\label{fig:molgas}
\end{figure}

In Figure~\ref{fig:molgas}$a$, we plot the molecular gas mass as a function of stellar mass for our high-z starbursts along with comparison samples (Section~\ref{sec:compsample}). Gas masses for the ALMA starbursts (filled red circles) range between 1.4 and $7.5\times10^{10}$ M$_{\odot}$, comparable or even exceeding their mass in stars, as indicated by the slanted solid line. As expected, these gas masses are substantially higher than low-redshift SF MS galaxies (grey circles) and starbursts (blue crosses). High-z SF MS galaxies (small triangles in purple) and the limited high-z starbursts samples (blue squares), with $\alpha_{CO}$ estimates, have similar values of M$_{gas}$/M$_{stellar}$ as our ALMA starbursts. 

In Figure~\ref{fig:molgas}$b$, we compare the ratio $\mu=M_{gas}/M_{stellar}$, relative to that of SF MS galaxies ($\mu_{MS}$) to our reference samples and the best-fit analytic expression given in \citet{Tacconi2018} as indicated by the green slanted line. The relative gas fraction is plotted as a function of the boost in sSFR relative to SF MS galaxies, as done in Figure~\ref{fig:tir_co}$b$. By comparing the ALMA starbursts (filled red circles) to the relation from \citet{Tacconi2018}, high-z starbursts have similar gas content to the more typical star-forming galaxies \citep{Daddi2010a,Tacconi2010}, counter to studies that favor higher gas fractions \citep{Genzel2015,Scoville2016,Lee2017}. We remark that this result is based on the implementation of $\alpha_{CO} = 1.3$ M$_{\odot}$ / (K km s$^{-1}$ pc$^{2}$). The use of a higher value of 4.36 M$_{\odot}$ / (K km s$^{-1}$ pc$^{2}$) for our high-z starbursts, as discussed further below, results in close agreement with the \citet{Tacconi2018} relation.

\subsection{Gas depletion times/star formation efficiency}

\begin{figure}
\epsscale{2.5}
\plottwo{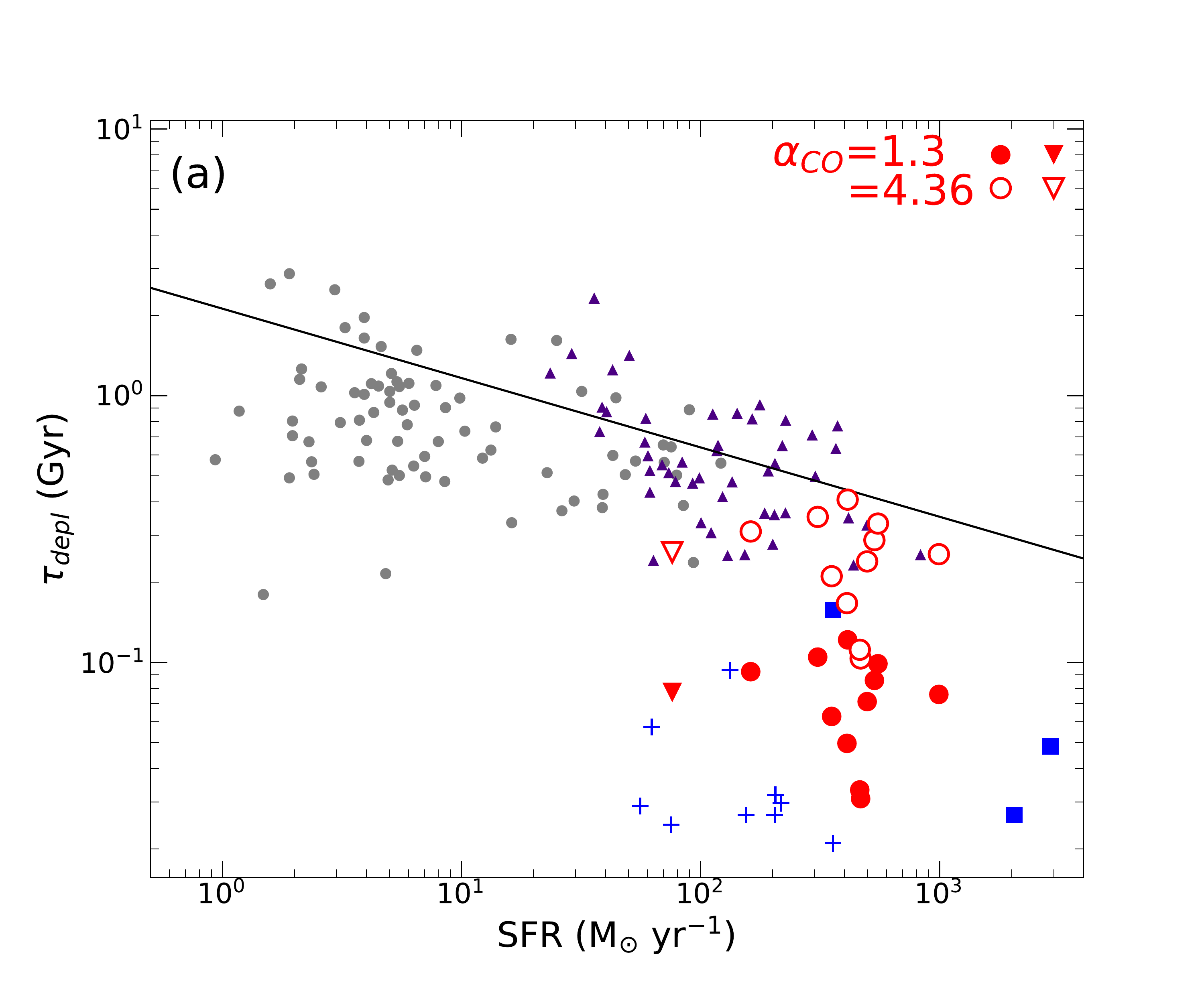}{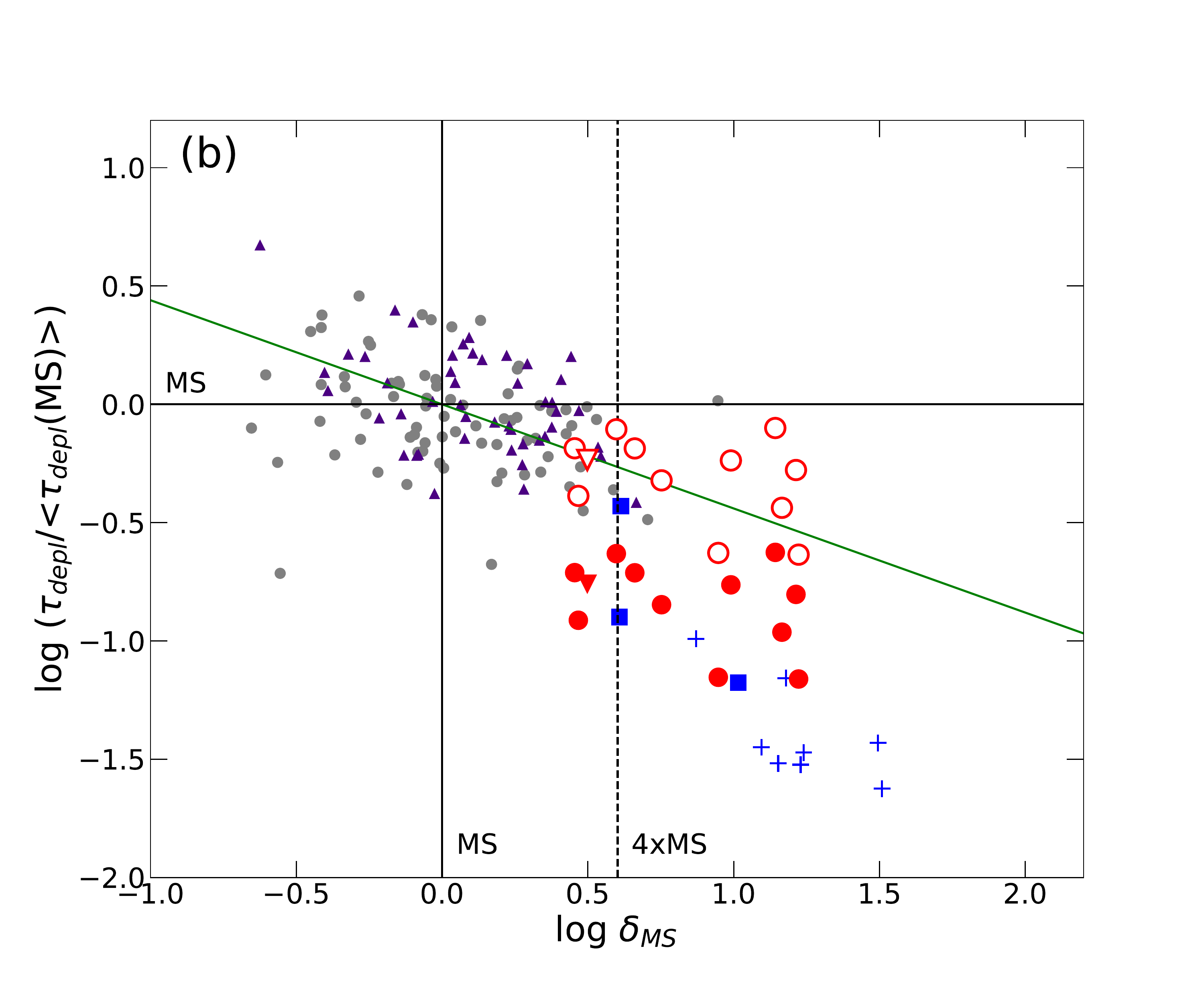}
\caption{(a) Gas depletion time ($\tau_{depl}=M_{gas}/SFR$) of high-redshift starbursts as a function of their SFR. Slanted black line is the best-fit relation from \citet{Sargent2014}. (b) Gas depletion time versus sSFR with both normalized to average values for SF MS galaxies. An analytic form of this relation is provided by \citet{Tacconi2018} and shown here in green. Symbols in both panels are the same as in Figure~\ref{fig:tir_co}.}
\label{fig:depl_time}
\end{figure}

With estimates of the gas mass, we can measure the efficiency of forming stars (SFE = SFR/M$_{gas}$) and its inverse, the time to deplete its gas reservoir if forming stars at a constant rate ($\tau_{depl}$=1/SFE) without gas replenishment. In Figure~\ref{fig:depl_time}$a$, we plot $\tau_{depl}$ as a function of SFR, comparing our high-z starbursts (filled red circles) to the reference samples described above. We find that our starburst sample has short gas depletion times ranging from $\sim40-100$ Myrs (Table~\ref{tab:physical}) that fall significantly offset from MS galaxies ($\sim0.4-1$ Gyrs) at a given SFR.

In Figure~\ref{fig:depl_time}$b$, the depletion times of our high-z starbursts drop even further from that of SF MS galaxies with increasing distance above the SF MS as quantified as $\delta_{\rm MS}$. The departure is still evident when comparing our data (red filled circles) to the analytical relation given in \citet{Tacconi2018} as indicated by the slanted green line. Based on these results, we reinforce our hypothesis put forward in \citet{Silverman2015a} that the star formation efficiency (or gas depletion time) increases (decreases) with distance above the MS stronger than any increase in the gas fraction, even if a mild increase is present.

\section{Discussion}
\label{text:discussion}

As with all CO studies of the molecular gas content, the essential inescapable issue is the conversion factor $\alpha_{\rm CO}$. As presented above, we have used a value of \hbox{1.3 M$_{\odot}$ / (K km s$^{-1}$} pc$^{2}$) based on a dynamical mass estimate from a higher resolution CO image of PACS-787 (Silverman et al. 2018, in press) that is broadly consistent with estimates from two other cases in our high-z starburst sample \citep{Silverman2015a}. This value is not too dissimilar to that used for local starbursts \citep[i.e., $\alpha_{CO}=0.8$ M$_{\odot}$ / (K km s$^{-1}$ pc$^{2}$);][]{Bolatto2013} and even actively star-forming regions of $z\sim2$ galaxies \citep[e.g.,][]{Tadaki2017}.

As mentioned above, our assessment of $\alpha_{\rm CO}$ relies on a $0\arcsec.3$ resolution map of PACS-787, acquired through a Cycle 4 ALMA program, that has revealed two galaxies in the process of merging with CO detections for each galaxy at a high signal-to-noise. While in an early stage of a merger, both galaxies have maintained their molecular gas disks thus enabling us to extract the gas mass from the dynamical mass with assumptions given in the aforementioned paper. It is possible that the value of the conversion factor used here induces a systematic level of uncertainty on the resulting gas masses. One issue may be that there is not definitive evidence for all starburst events within our sample as being triggered by a major merger, particularly at the same stage as PACS-787 in an evolutionary merger sequence. However, the majority of starbursts in our ALMA sample do have multiple emitting components in the rest-frame ultraviolet (e.g., PACS-819, 830, 491) and optical (e.g., PACS-867, 299, 325, 164) as detected by $HST/ACS$ and $Spitzer$ \citep[see Figure 2 of][]{Silverman2015a} thus we presume that most of our starbursts are undergoing some sort of an interaction or merging event. Still, there is always the possibility that the conversion factor may be larger. Therefore, we explore the impact of a higher conversion factor on our results. \citet{Tacconi2018} are advocating the application of a more universal value ($\alpha_{CO}=4.36$ M$_{\odot}$ / (K km s$^{-1}$ pc$^{2}$), plus a metallicity dependence) for the analysis of large samples. 

First, we assess the impact on the gas fractions and depletion times if the CO-to-H$_2$ conversion factor is 4.36 M$_{\odot}$ / (K km s$^{-1}$ pc$^{2}$). This value is similar to star-forming regions in our own Milky Way and nearby local galaxies \citep{Bolatto2013}. As a result, the gas fractions, expressed here as $M_{gas}/ M_{stellar}$, will be appreciably higher with most above unity since they will be dominated in their mass budget by the molecular gas. This is seen in Figure~\ref{fig:molgas}$a$ where gas mass is plotted as a function of the stellar mass. In Figure~\ref{fig:molgas}$b$, the ratio of the gas mass to stellar mass is plotted as a function of $\delta_{\rm MS}$ as defined above. The new gas masses (open red circles) are found to be in closer agreement to the relation from \citet{Tacconi2018} that shows an increase in gas mass fraction, as compared to SF MS galaxies, with their boost in sSFR above the MS. In this scenario, the boost in SFR for starbursts would be attributed to both an increase in the gas fraction and the SFE. This is seen in Figure~\ref{fig:depl_time}$b$ where the depletion times are significantly shorter than SF MS galaxies for $\alpha_{\rm CO}=1.3$ M$_{\odot}$ / (K km s$^{-1}$ pc$^{2}$), while a higher CO-to-H$_2$ conversion factor results in depletion times that are relatively higher but still slightly shorter than MS galaxies at all redshifts.

In general, any application of a scale factor to a set of measurements based on a well-selected statistical sample should be based on an independent assessment of the validity of this factor using a representative subset of the sample. To achieve such assurance in the applicable value of $\alpha_{\rm CO}$ to high-z starbursts (selected as outliers above the star-forming MS), we have obtained ALMA observations at a spatial resolution of $\sim1\arcsec$ or below for three starbursts (\citealt{Silverman2015a}; Silverman et al. 2018, in preparation) that provide broadly consistent results for a low value of $\alpha_{\rm CO}$. Although, we note that the normalization of the dynamical mass estimate and the method to assess the errors differ between PACS-819/830 and PACS-787. For both cases, there remains uncertainty in these estimates and further effort with ALMA is needed at higher resolution to improve the quality of the size measurements, inclinations, and presence of any inflow/outflow components to the CO emission. 

In light of these uncertainties, we are confident that the main results of this study are robust when considering the observed quantities irrespective of the value of $\alpha_{\rm CO}$. In Figure~\ref{fig:tir_co}a, it is clear that the high-z starbursts are offset from the relation between $L^{\prime}_{CO}$ and $L_{\rm TIR}$ for MS galaxies. There appears to be a continuous decline in the amount of CO-emitting gas with distance above the MS (Fig.~\ref{fig:tir_co}b) thus supporting models of such behavior \citep{Narayanan2012}. 

\label{sec:discussion}

\section{Final Remarks}

The question of whether starburst outliers from the MS owe their higher sSFR to a higher gas fraction or to a higher star formation efficiency (or combination thereof) remains an unsettled issue in the current literature. For example, \citet{Sargent2014} and \citet{Silverman2015a} argue for a substantially higher SFE, with even sub-MS gas fractions, whereas \citet{Scoville2016} support the opposite notion of starbursts being primarily driven by a higher gas fraction in outliers compared to MS galaxies. More recently, scaling relations have been proposed that tend to combine both effects. Thus, \citet{Scoville2017} derive scaling relations from their ALMA dust continuum observation according to which

\begin{equation}
{{\rm SFR}\over M_{\rm gas}}={\rm SFE} \propto \delta_{\rm MS}^{0.70}
\end{equation}
and
\begin{equation}
\mu={M_{\rm gas}\over M_{\rm stellar}}\propto \delta_{\rm MS}^{0.32}M_{\rm stellar}^{-0.7}
\end{equation}

\noindent respectively from Eq. (7) and (6) of \citet{Scoville2017}. Therefore, the elevation of the sSFR over the MS ($ \delta_{\rm MS}$), at fixed stellar mass, is attributed  to a combination of both a higher gas fraction and a higher SFE, scaling as $\sim \mu^3$ and $\sim$SFE$^{1.4}$ respectively. Given that the gas fraction is higher in MS lower mass galaxies (scaling as  $M_{\rm stellar}^{-0.7}$), upon a major merger a galaxy would find itself with a higher gas fraction compared to a MS galaxy with a stellar mass equal to the combined stellar mass of the merger, which according to Scoville et al. explains the higher gas fraction that they derive for MS outliers.

The equivalent scaling relations given by \citet{Tacconi2018} using primarily CO data  are somewhat different:

\begin{equation}
{{\rm SFR}\over M_{\rm gas}}={\rm SFE} \propto \delta_{\rm MS}^{0.44}
\end{equation}

and

\begin{equation}
\mu={M_{\rm gas}\over M_{\rm stellar}}\propto \delta_{\rm MS}^{0.53}M_{\rm stellar}^{-0.35}
\end{equation}

\noindent Hence, for starbursting outliers, they find a slightly lower dependence of $\delta_{MS}$ on SFE, as compared to \citet{Scoville2017}, along with a slightly higher dependence on gas fraction, compared to MS galaxies of the same stellar mass. The higher gas fraction (a factor of $\sim 3$ for extreme starburst with $ \delta_{\rm MS}=10$) could not be completely accounted by simple merging, as above, because of the flatter dependence of $\mu$ on the stellar mass along the MS, as from Equations 3 and 5. To account for this higher (molecular) gas fraction, one may invoke some conversion of HI to H$_2$, as the circumgalactic HI reservoir could be destabilized upon merging. Note that \citet{Tacconi2018} adopt a universal $\alpha_{\rm CO}=4.36$ M$_{\odot}$ / (K km s$^{-1}$ pc$^{2}$) (scaled down with increasing metallicity), for both MS and MS outliers.

As shown in Figure~\ref{fig:molgas}$b$, adopting our best fit value from the dynamical argument $\alpha_{\rm CO}=1.3$ M$_{\odot}$ / (K km s$^{-1}$ pc$^{2}$) results in gas fractions for starburst galaxies that are on average similar to MS galaxies, hence the elevation of sSFR is fully attributed to a higher SFE, with SFE$\sim\delta_{MS}$, and in extreme starbursters the SFE should be up to $\sim 30$ times higher than on the MS. Sargent et al. (2014) argue that a lower gas fraction in starbursts would be a natural result of their much shorter gas depletion time $\equiv {\rm SFE}^{-1}$, if on average they are caught midway through their starburst, having already consumed a
major fraction of their gas reservoir. This implies that starbursts at high redshifts would nearly double their stellar mass during the starburst, which at first sight may conflict with the constraint according to which on a global scale starbursts contribute for just $\sim 15\%$ of star formation (Rodighiero et al. 2011). However, one may argue that only a fraction, say, $\sim 1/3$, of massive galaxies experience a merger driven starbursts.

Adopting $\alpha_{\rm CO}=4.36$ M$_{\odot}$ / (K km s$^{-1}$ pc$^{2}$) as shown in Figure~\ref{fig:molgas}$b$, the starbursts fall along the continuous extension from the MS thus indicative of a single mode of star formation, whereas departures from such relations are more evident when adopting different values of $\alpha_{\rm CO}$ for MS and starburst galaxies. These examples illustrate how much divergence there is from the resulting interpretations of why starburst galaxies have elevated sSFRs, with a major role being played by the {\it infamous} $\alpha_{\rm CO}$.

\acknowledgments

We are grateful for the support from the regional ALMA ARCs. JDS was supported by the ALMA Japan Research Grant of NAOJ Chile Observatory, NAOJ-ALMA-0127.  This work was supported by World Premier International Research Center Initiative (WPI Initiative), MEXT, Japan. CM and AR acknowledge support from an INAF PRIN 2012 grant.  W.R. is supported by Thailand Research Fund/Office of the Higher Education Commission Grant Number MRG6080294 and Chulalongkorn University's CUniverse. GEM acknowledges support from the Carlsberg Foundation, the ERC Consolidator Grant funding scheme (project ConTExt, grant num- ber No. 648179), and a research grant (13160) from Villum Fonden. N.A. is supported by the Brain Pool Program, which is funded by the Ministry of Science and ICT through the National Research Foundation of Korea (2018H1D3A2000902). This paper makes use of the following ALMA data: ADS/JAO.ALMA\#2012.1.00952.S, ADS/JAO.ALMA\#2015.1.00861.S and ADS/JAO.ALMA\#2016.1.01426.S. ALMA is a partnership of ESO (representing its member states), NSF (USA) and NINS (Japan), together with NRC (Canada), NSC and ASIAA (Taiwan), and KASI (Republic of Korea), in cooperation with the Republic of Chile. The Joint ALMA Observatory is operated by ESO, AUI/NRAO and NAOJ.

\begin{deluxetable*}{lllllllll}
\tabletypesize{\scriptsize}
\tablecaption{Starburst sample\tablenotemark{a} \label{tab:sample}}
\tablehead{\colhead{ID}&\colhead{RA}&\colhead{Dec}&\colhead{$z_{spec}$\tablenotemark{b}}&\colhead{log M$_{stellar}$\tablenotemark{c}}&\colhead{$L^{Total}_{IR}$}&\colhead{SFR (IR)}&\colhead{log $\delta_{\rm MS}$\tablenotemark{d}}\\
&&&&\colhead{(M$_{\odot}$)}&\colhead{(L$_{\odot}$)}&\colhead{(M$_{\odot}$ yr$^{-1}$})}
\startdata
787&10:02:27.95&02:10:04.4&1.5234&10.56&12.83$\pm$0.04&991$^{+96}_{-87}$&0.8\\
197&10:01:34.46&01:58:47.7&1.6005&10.75&12.58$\pm0.06$&551$^{+83}_{-72}$&0.7\\
491&10:00:05.16&02:42:04.7&1.6366&10.35&12.50$\pm0.08$&463$^{+96}_{-80}$&0.9\\
224&09:58:56.51&02:03:47.5&1.6826&10.05&12.51$\pm0.10$&467$^{+121}_{-96}$&1.2\\
251&10:02:39.63&02:08:47.2&1.5847&10.05&12.45$\pm0.10$&409$^{+101}_{-81}$&1.2\\
837&10:00:36.31&02:21:17.5&1.6552&9.98&11.72$\pm0.12$&76$^{+25}_{-19}$&0.5\\
\hline
299&09:59:41.31&02:14:42.8&1.6467&10.09&12.53$\pm0.15$&$497^{+208}_{-147}$&1.2\\
325&10:00:05.53&02:19:42.83&1.6557&10.39&12.05$\pm0.22$&162$^{+106}_{-64}$&0.4\\
819&09:59:55.54&02:15:11.46&1.4449&10.37&12.56$\pm0.05$&533$^{+68}_{-60}$&1.0\\
830&10:00:08.73&02:19:02.47&1.4610&10.68&12.45$\pm0.06$&412$^{+62}_{-54}$&0.6\\
867&09:59:38.10&02:28:57.06&1.5673&10.75&12.38$\pm0.18$&353$^{+179}_{-119}$&0.5\\
164&10:01:30.42&01:54:12.50&1.6489&9.94&12.33$\pm0.28$&309$^{+274}_{-145}$&1.1\\
\enddata
\tablenotetext{a}{The horizontal line differentiates between galaxies observed in Cycles 1 (below) and 3 (above) with the exception of PACS-164 that was observed with NOEMA.}
\tablenotetext{b}{Spectroscopic redshifts are based on H$\alpha$ and have errors $\sigma_{\Delta z/(1+z)}=1.8\times10^{-4}$.}
\tablenotetext{c}{$\sigma_{\rm M}\sim0.07$ dex error on the stellar mass \citep{Ilbert2015}}
\tablenotetext{d}{$\delta_{\rm MS}$ = sSFR/$<sSFR_{MS}>$}
\end{deluxetable*}

\begin{deluxetable*}{lllllllllll}
\tabletypesize{\scriptsize}
\tablecaption{ALMA Cycle 3 CO(2 - 1) measurements\label{tab:co}}
\tablehead{\colhead{ID}&\colhead{RA}&\colhead{Dec}&\colhead{$z_{CO}$\tablenotemark{a}}&\colhead{I$_{CO}$\tablenotemark{b}}&\colhead{$\Delta$v\tablenotemark{c}}&\colhead{$L_{CO}^{\prime}$\tablenotemark{d}}&\colhead{Beam}&\colhead{$\sigma_{rms}$\tablenotemark{f}}\\
&\colhead{(CO)}&\colhead{(CO)}&&&&&\colhead{size\tablenotemark{e}}}
\startdata
787&10:02:27.954&+02:10:04.40&1.5249&1.64$\pm$0.20&600&10.69&2.67$\times$1.79 (67.0)&0.085\\
197&10:01:34.461&+01:58:47.69&1.6016&1.086$\pm$0.088&700&10.55&2.11$\times$1.79 (-69.6)&0.117\\
491&10:00:05.164&+02:42:04.69&1.6358&0.295$\pm$0.038&300&10.00&2.16$\times$1.73 (-69.0)&0.067\\
224&09:58:56.515&+02:03:47.70&1.6826&0.262$\pm$0.034&400&9.98&2.39$\times$2.02 (63.3)&0.026\\
251&10:02:39.626&+02:08:47.19&1.5863&0.413$\pm$0.040&700&10.12&2.36$\times$2.01 (62.4)&0.040\\
837&10:00:36.233&+02:21:18.08&1.6572&0.127$\pm$0.061&300&$<$9.66&2.38$\times$2.02 (62.8)&0.026\\
\enddata
\tablenotetext{a}{$\sigma_z$ = 0.0003-0.0004.}
\tablenotetext{b}{units of Jy km s$^{-1}$}
\tablenotetext{c}{Full width of the CO line at zero intensity in units of velocity (km s$^{-1}$).}
\tablenotetext{d}{{\rm Log} base 10; units of K km s$^{-1}$ pc$^2$}
\tablenotetext{e}{units of arcsecs (position angle in degrees).}
\tablenotetext{f}{units of Jy beam$^{-1}$ km s$^{-1}$}
\end{deluxetable*}

\begin{deluxetable}{llll}
\tabletypesize{\scriptsize}
\tablecaption{Derived physical properties\label{tab:physical}}
\tablehead{\colhead{ID}&\colhead{log $M_{gas}$\tablenotemark{a}}&\colhead{$M_{gas}/M_{stellar}$}&\colhead{$\tau_{depl}$\tablenotemark{b}}}
\startdata
787&10.88&0.9&76\\
197&10.74&1.0&99\\
491&10.19&0.7&33\\
224&10.16&1.3&31\\
251&10.31&1.8&50\\
837&$<$9.77&$<$0.6&$<$77\\
\hline
299&10.55&2.9&71\\
325&10.17&0.6&92\\
819&10.66&1.9&86\\
830&10.70&1.0&121\\
867&10.34&0.4&63\\
164&10.51&3.7&105\\
\enddata
\tablenotetext{a}{units of M$_{\odot}$ and based on $\alpha_{CO}=1.3$}
\tablenotetext{b}{$\tau_{depl}=M_{gas}/SFR$; units Myr}
\end{deluxetable}

\appendix

\section{Location of CO-emitting regions}

In \citet{Silverman2015a}, we reported that the CO sources were usually offset from regions of rest-frame UV emission, as seen by the Hubble Space Telescope (HST) Advanced Camera for Surveys (ACS) observations of COSMOS \citep{Scoville2007,Koekemoer2007}. However, the centroid of the CO emission was always co-spatial with the stellar mass distribution as traced by the $Spitzer$/IRAC 3.6 - 8 $\mu$m bands. To investigate the significance of CO/UV offsets in more detail, we use recently reprocessed HST/ACS mosaics (Koekemoer 2018, priv. comm) where the astrometry is directly tied to the International Celestial Reference Frame (ICRF), thereby eliminating uncertainties from possible offsets in older catalogs previously used to align HST images. The astrometric comparisons with our new ALMA observations corroborate the significance of CO/UV offsets found in our past results, as shown in Figure~\ref{fig:co_images}. For example, PACS 251, 787 and 491 have UV emission (shown by the grey scale image) offset from the CO emission (blue contours) while more closely associated with the IR emission (red contours). In fact, the other two (PACS 197 and 224) have essentially no emission in the UV while strongly detected by $Spitzer$. These results are in agreement with our recent studies \citep{Silverman2015a,Puglisi2017} and many others on the extremely dusty nature of starbursts \citep[e.g., ][]{Smail2004,Ivison2010,Chen2015} where any UV or rest-frame optical emission lines are emitted from regions of relatively low extinction, possibly close to the surface of embedded star-forming regions since the bulk ($\sim90\%$) of the star formation is only observed at infrared wavelengths or longer.

\end{document}